\pdfoutput=1
\documentclass[a4paper,12pt]{article}
\usepackage{natbib,epsfig, graphics, lscape, longtable,
rotating, geometry, threeparttable,amsmath,multirow,tabularx,setspace,amssymb}

\usepackage{enumerate}

\usepackage{amsmath,mathtools,amsthm,float}

\usepackage{enumitem}

\usepackage{titlesec}

\usepackage{appendix}
\usepackage[bottom]{footmisc}

\usepackage{graphicx}
\usepackage[centerlast]{caption}
\usepackage{subcaption}
\usepackage{placeins}

\usepackage{booktabs}

\usepackage{pdflscape}
\usepackage{fancyhdr}
\usepackage{adjustbox}
\usepackage[absolute,overlay]{textpos}

\usepackage{hyperref}
\hypersetup{
    colorlinks=true, %
    linkcolor=black, %
    citecolor=black, %
    filecolor=black,
    urlcolor=blue
}

\usepackage{tikz}
\usetikzlibrary{calc,arrows,automata,shapes.misc,shapes.arrows,chains,matrix,positioning,scopes,decorations.pathmorphing,shadows,shapes,snakes,fadings}
\usetikzlibrary{chains, positioning, shapes.symbols}
\tikzset{start/.style = {signal, fill=#1,minimum height=12mm,draw=white,
                 font=\footnotesize,text=black,inner sep=4pt,
                 signal pointer angle=150, on chain},
  cont/.style = {start=#1, signal from=west},
  end/.style = {start=#1, signal from=west, signal pointer angle=150}}
\tikzset{
    invisible/.style={opacity=.5},
    visible on/.style={alt={#1{}{invisible}}},
    highlight on/.style={alt={#1{fill=title.fg!25!white}{fill=gray!20!white}}},
    alt/.code args={<#1>#2#3}{%
      \alt<#1>{\pgfkeysalso{#2}}{\pgfkeysalso{#3}}
    },
  }

\newtheorem{definition}{Definition}

\newtheorem{assumption}{Assumption}

\newtheorem{result}{Result}

\renewcommand{\footnotesize}{\small}

\renewcommand{\thesection}{\Roman{section}}
\renewcommand{\thesubsection}{\thesection.\Alph{subsection}}

\usepackage{dsfont}
\usepackage{threeparttable}

\usepackage{changepage}

\begin{document}

\title{\vspace{-1cm} Identifying level-$k$ reasoning in repeated games:\\ Strategies, beliefs, and cognitive ability
\vspace{0.5cm}
\thanks{We thank: Shabnam Rahimfallah, Rui Tan and Zhongheng Qiao for excellent
research assistance; seminar participants and colleagues for valuable comments and discussion;
the Vernon Smith Experimental Economics Laboratory at Purdue University for hosting our
experiment; and the Blake Family Fund for Ethics, Leadership and Governance (administered by
the Daniels School of Business) for providing funding. This paper supersedes Section 3.3 of \cite{GilRos_20}, which contained a preliminary version of some of the analysis here, and which was not included or referenced by \cite{GilRos_24}.}
 }

\date{}

\author{
\large{David Gill \thanks{Purdue University, Department of Economics; dgill.econ@gmail.com}}\\
\large{Yaroslav Rosokha \thanks{Purdue University, Department of Economics; yrosokha@purdue.edu}}\\
}

\maketitle

\thispagestyle{empty}

\vspace{-0.5cm}
\begin{center}
This version:  October 22, 2026\\
First version: January 19, 2026\\

\end{center}

\vspace{0.1cm}

\begin{abstract}
\begin{adjustwidth}{-2mm}{-2mm} %
\setstretch{1.3}
\noindent In this paper, we identify level-$k$ reasoning in repeated games that operates at the level of a supergame strategy. First, we develop a model of level-$k$ reasoning that incorporates choices over strategies as well as beliefs about the strategies chosen by others. Then, jointly using elicited strategies and beliefs about strategies from the indefinitely repeated Prisoner's Dilemma, and applying a new notion of balance-optimal error tolerance, we classify a substantial fraction of subjects as level-1 or level-2. Moreover, we show that when level-$k$ reasoning operates at the level of a repeated-game strategy, cognitive ability and experience predict higher level reasoning.
\vspace{1cm}

\textbf{Keywords}: \textit{Level-$k$, repeated game, Prisoner's Dilemma, strategy, beliefs, cognitive ability, experience, elicitation, bounded rationality, experiment, game theory.}\\

\textbf{JEL Classification}: \textit{C73, D83, D91.}
\end{adjustwidth}
\end{abstract}

\newpage

\newpage
\pagenumbering{arabic}

\section{Introduction}\label{Sect:Intro}

The level-$k$ model of boundedly rational reasoning is a powerful tool for analyzing strategic interactions. The model eschews equilibrium thinking; instead, level-$k$ reasoning anchors beliefs in an instinctive reaction to the game and then adjusts them via iterated best responses \citep{CraCosIri_13}. Specifically, the level-0 type is strategically unsophisticated, the level-1 type best responds to the belief that all others are level-0, the level-2 type best responds to the belief that all others are level-1, and so on \citep{Nag_95}. \cite{AlaPen_16, AlaPen_22}’s model of endogenous depth of reasoning microfounds choice of level, while \cite{AloBuc_21} and \cite{GilPro_23} find that higher level-$k$ types think for longer. The level-$k$ model has been used to help understand strategic behavior in a variety of settings, including auctions (e.g., \citealp{CraIri_07}), undercutting games (e.g., \citealp{AraRub_12}), Centipede games (e.g., \citealp{GarIriKov_20}), Cournot games (e.g., \citealp{HoSu_21}), and gas station pricing (\citealp{HanHufLiu_26}).

Despite the central role that beliefs play in level-$k$ theory, empirical applications have not used choices and elicited beliefs together to estimate models of level-$k$ behavior.\footnote{There are two main reasons for this. First, the existing literature often studies relatively simple games in which different levels generate distinct choice predictions, making it possible to infer levels from choices alone. Second, accommodating noise in both beliefs and choices, and then jointly mapping these two sources of data into levels of reasoning, is methodologically challenging.} In this paper, we use choice and belief data jointly to identify level-$k$ reasoning in the indefinitely repeated Prisoner's Dilemma (IRPD), which captures well the trade-off between the short-term payoff from exploiting partners and the long-term gain from building enduring cooperative relationships. Strategies in the IRPD incorporate various forms of sophisticated behavior, including conditional cooperation, punishment, and forgiveness. Given this complexity, we show that data on beliefs are critical for level-$k$ identification because strategy choices alone cannot clearly distinguish different levels, or indeed level-$k$ reasoning from other models of behavior.

We measure level-$k$ reasoning that operates at the level of a supergame strategy (i.e., a plan of action for every round of the repeated game that can condition on the history of play). By contrast, the existing literature that measures level-$k$ reasoning in repeated games models level-$k$ behavior at the level of a single round or repetition.\footnote{Some papers estimate level-$k$ reasoning separately in each round (\citealp{Nag_95}, \citealp{DufNag_97}, \citealp{DanFefKub_12}, \citealp{LinSut_13}, \citealp{BenKubNor_17}, \citealp{FerPitGar_25}). Others use data from multiple rounds to estimate level-$k$ reasoning (\citealp{Sta_96}, \citealp{HoCamWei_98}, \citealp{GilPro_16}, \citealp{HoSu_21}, \citealp{CasProSof_23}, \citealp{GilPro_23}, \citealp{GilProKne_25}), but the level-$k$ reasoning remains defined at the level of stage-game actions (even when beliefs adjust dynamically, e.g., about the behavior of level-0 types as in  \citealp{GilPro_16}).  Given this focus, the literature has not measured level-$k$ reasoning in settings such as the IRPD, where round-by-round level-$k$ analysis is silent about depth of reasoning because all types with $k \geq 1$ defect in the stage game.} We follow a recent stream of papers that directly elicits strategies in the IRPD (\citealp{RosRom_18, RomRos_23}, \citealp{CasMui_19}, \citealp{DalFre_19}). Specifically, we use the IRPD dataset from our earlier study, \cite{GilRos_24}.\footnote{Although we use the dataset from \cite{GilRos_24}, we pursue different empirical goals. \cite{GilRos_24} do not structurally estimate types; instead, using only reduced-form analysis, the paper focuses on the relationship between cooperation and beliefs about cooperation, and how both evolve with experience.  \cite{GilRos_24} find that being matched with a cooperative opponent increases subsequent cooperation and makes beliefs more optimistic. In this paper, we do not study individual-level responses to experience; instead, we use strategies and beliefs about strategies to identify level-$k$ reasoning, and we study how level-$k$ reasoning changes with experience at the aggregate level.} In that experiment we elicit supergame strategies and beliefs about strategies: subjects choose a strategy at the beginning of each supergame, rather than playing via direct response (with a sequence of round-by-round decisions), and report beliefs over the strategies chosen by others (in the first and final supergames).

Our dataset is particularly well suited to identifying level-$k$ reasoning that operates at the level of a strategy because it contains both strategy choices and beliefs. Access to data on beliefs is essential for identification, for two reasons. First, different level-$k$ types may choose the same strategy while holding different beliefs. For example, two subjects may choose Always Defect (AD), yet one may do so because she believes opponents randomize uniformly across strategies (holding level-1 beliefs), while the other may do so because she believes opponents will choose AD (holding level-2 beliefs). In this case, observing beliefs allows us to distinguish between reasoning types despite identical strategy choices. Second, the same strategy choice could arise from level-$k$ reasoning or from some other process (e.g., confusion, imitation, efficiency seeking, equilibrium thinking). In this case, observing beliefs allows us to assess whether a subject is consistent with the level-$k$ model. That is, beliefs help identify the reasoning behind chosen strategies, thereby strengthening identification in a belief-driven model like level-$k$.  In Section \ref{Sect:Estimation}, we develop these points further, using specific experimental subjects as examples.

We use elicited strategies and beliefs about strategies instead of direct-response within-supergame round-by-round choices and beliefs. Strategy elicitation is a well-established experimental design that goes back to \cite{Sel_67} and comes with benefits and costs. In our case, the main benefit is that we directly observe strategies and beliefs about strategies. If we used round-by-round data, we would need to structurally estimate the distributions of strategies and beliefs \citep{AoyFreYuk_24}, and then use posterior probabilities to noisily allocate specific strategies and beliefs to each subject. Furthermore, we study how experience affects level-$k$ reasoning: structural estimation requires the assumption of stable strategy choices and beliefs across many supergames, and so round-by-round data cannot be used to measure level-$k$ reasoning that operates at the level of a strategy before subjects gain experience. Another benefit is that our subjects evaluate strategies only at the beginning of the supergame, as assumed in the theoretical analysis of repeated games.

The main cost is that strategy elicitation could lead to behavior that differs from direct-response play. However, previous evidence and features of our setting help to allay this concern. First, existing evidence shows that when strategies are elicited in the IRPD, the empirical distribution of strategies is similar to that estimated from direct-response play (\citealp{RosRom_18}, \citealp{DalFre_19}).\footnote{To enable elicitation of both strategies and beliefs, we restrict attention to 10 strategies. While such a restriction could in principle distort behavior, these strategies include the 8 most common identified via structural estimation using round-by-round choices in the IRPD (\citealp[Section I.D]{GilRos_24}). Consistent with this, our data from elicited strategies replicate key features of direct-response play, although cooperation is broadly stable over supergames when the return to cooperation is high (see \citealp[Appendix VII]{GilRos_24}, for details).}  This is consistent with survey evidence in \citet{BraCha_11} from more general settings (they find no case where the strategy method generates new treatment effects, and that in a majority of surveyed studies behavior is similar under the strategy method and direct response). Second, we emphasize that strategy elicitation captures the dynamic reasoning at the core of repeated-game play. The object of choice remains a repeated-game strategy (a history-contingent plan), with strategies varying in whether and how quickly they punish and relent (see Figure \ref{Fig:Strategies}). Thus, the decision problem involves reasoning over the full dynamic paths of play implied by pairs of strategies. As in direct-response play, subjects need to consider whether others are likely to start by cooperating or defecting, how harshly defections are likely to be punished, and whether attempting to return to cooperation after a period of punishment is worthwhile. Third, our design keeps the round-by-round representation of the repeated game salient throughout. Before choosing strategies, subjects gained experience with direct-response play in practice supergames played against themselves, and once strategies were chosen, every supergame was played out round-by-round on the subject's screen similarly to how direct-response games unfold. Fourth, given our experimental parameters, full cooperation (via Grim) and full defection (via Always Defect) are both Nash equilibria of our game with strategy elicitation and subgame-perfect Nash equilibria (SPE) of the direct-response IRPD. Thus we preserve equilibrium multiplicity.\footnote{Formally, strategy elicitation induces a one-shot normal-form game whose set of Nash equilibria does not coincide exactly with the set of SPE of the direct-response IRPD (differences are due to the number of strategies and sequential rationality).}

Our first contribution is to develop a model of level-$k$ reasoning in repeated games that operates at the level of a supergame strategy. The model incorporates choices over strategies as well as beliefs about the strategies chosen by others. The strategically unsophisticated level-0 type randomizes uniformly over the set of strategies. The level-1 type holds beliefs within distance $\eta$ of uniform and chooses a strategy within distance $\eta$ of a best response to her beliefs, where $\eta \geq 0$ indexes the degree of error tolerance in terms of distance. The level-2 type holds beliefs within distance $\eta$  of a distribution consistent with level-1 behavior, and again chooses a strategy within distance $\eta$  of a best response to her beliefs. We measure distance in payoff space, defining distance for both choices and beliefs as proportional payoff loss, with belief payoffs given by the Quadratic Scoring Rule used in our experiment. We do not include higher level types in our estimation, but the model can be extended to include analogous higher level types.

Our second contribution is to show that our model of level-$k$ reasoning can help to understand observed strategy choices and beliefs in repeated games. Specifically, using the IRPD dataset from \cite{GilRos_24}, we classify each subject according to whether her strategy choices and her elicited beliefs are jointly consistent with the level-1 type, the level-2 type, both types, or neither type, at a given level of error tolerance. Classification therefore requires choosing the level of error tolerance, and this choice involves a trade-off. Tolerating too little error means that few subjects conform to the level-$k$ model, while tolerating too much error means that the behavior of the level-1 and level-2 types overlaps excessively. Limiting error tolerance also guards against classifying as level-1 or level-2 those subjects whose behavior in fact arises from some other process. We resolve this trade-off by introducing the new notion of \textit{balance-optimal error tolerance}, defined as the tolerance that maximizes the share of subjects classified uniquely as level-1 or level-2. At the balance-optimal error tolerance, we find that a substantial fraction of subjects are classified as level-1 or level-2 types. Based on the type classification, we also investigate how level-$k$ types map empirically to beliefs and strategy choices.

We are not aware of any prior paper that measures level-$k$ reasoning that operates at the level of a supergame strategy in a repeated game. \cite{WunLitSto_09}'s purely theoretical exercise calculates higher-order best responses for a cognitive hierarchy model defined over strategy choice in the finitely repeated Prisoner's Dilemma. \citet[Section 5.1]{CasPro_20} note that the positive relationship between anger and AD documented by \cite{CasMasPro_18} is consistent with anger increasing level-0 behavior, because a level-$k$ model where level-0 uniformly randomizes over \{AD, G, TFT\} predicts that only the level-0 type chooses AD in their IRPD. Finally, \citet[Section V]{AoyFreYuk_24} use numerical computation based on exogenously given parameters to show that a cognitive hierarchy model defined over strategy choice can generate patterns consistent with their repeated Prisoner's Dilemma data; however, they do not estimate proportions of level-$k$ types (nor do they use data on beliefs in this part of the paper).\footnote{A small literature measures level-$k$ reasoning that operates at the level of a supergame strategy in the sequential Centipede game. \cite{GarIriKov_20} use elicited strategies but not beliefs, while  \cite{KawTak_12} and \cite{HoSu_13} elicit neither strategies nor beliefs. See also \cite{SchZho_24}, who develop their model of strong level-$k$ thinking for extensive-form games, and \cite{LinPal_24}, who develop a dynamic cognitive hierarchy model for extensive-form games.}$^{,}$\footnote{\cite{HanHufLiu_26} find that high-cognitive-skill gas station managers set higher prices. Using level-$k$ estimated from a one-shot  game (a variant of the 11-20 game), and using responses to unincentivized survey questions, they interpret gas station manager behavior through the lens of a level-$k$ model of repeated pricing in which level-2 managers anticipate responses of rivals to price cuts, while level-1 managers do not.}

Furthermore, we are not aware of any prior level-$k$ paper that elicits beliefs over strategies (in repeated or sequential games); nor are we aware of any paper that jointly uses elicited beliefs and choices to estimate the distribution of level-$k$ types. In the context of repeated games with level-$k$ estimated round by round, \cite{DanFefKub_12} estimate the distribution of level-$k$ using only elicited beliefs, and separately using only choices, finding a high degree of within-individual consistency in estimated levels. In one-shot games, \cite{HynTerVak_22} calculate the frequencies of different levels, averaged across games, again using elicited beliefs and choices separately.\footnote{Other papers that use elicited beliefs in one-shot games in the context of level-$k$ include \cite{Har_02}, \cite{CosWei_08}, \cite{Rey_09}, \cite{HynOzbSch_12}, \cite{BurPen_14}, \cite{Lah_15}, \cite{GoeLouZha_18}, \cite{PolCor_19}, \cite{FraKovRoj_20}, \cite{AleColKol_22}, \cite{GoeGar_23}, and \cite{FriWar_25}.}

A growing literature studies the role of cognitive ability in strategic behavior (see the review by \citealp{Sof_25}). Cognitive ability supports logical reasoning and the organization of new information. In strategic settings, higher cognitive ability enhances strategic skill by improving individuals’ understanding of the strategic environment, which in turn facilitates the formation of beliefs about others and supports better choices given those beliefs \citep{FeGilPro_22}. We measured cognitive ability using matrix reasoning, which captures fluid intelligence (i.e., logical or analytical intelligence) and is a leading measure of cognitive ability that correlates strongly with general intelligence; our test is similar to Raven’s Progressive Matrices (see Section \ref{Sect:Design} for details).

Our third contribution is to provide evidence that cognitive ability predicts higher level-$k$ reasoning when level-$k$ reasoning operates at the level of a supergame strategy. Specifically, using our IRPD dataset, we show that subjects with higher cognitive ability are more likely to be classified as level-2 rather than level-1. Our findings complement \cite{ProRusSof_19,ProRusSof_22}, who find that cognitive ability influences cooperation in the IRPD because more cognitively able subjects make fewer errors when implementing their repeated-game strategies. Our design with strategy elicitation shuts down their primary mechanism.\footnote{Consistent with this, \citet[Appendix VIII]{GilRos_24} report that cognitive ability does not predict cooperation in our data. \citet[Appendix VIII]{GilRos_24} also report that cognitive ability modestly predicts belief accuracy in the final supergame.} Instead, we find a novel channel through which cognitive ability influences behavior in the IRPD, namely through higher level reasoning in belief formation and strategy selection. The IRPD captures well complex strategy choice in repeated games that allow cooperation, punishment and forgiveness. To the extent that our findings extend to more general repeated games, accounting for higher level-$k$ reasoning will sharpen predictions about how cognitive ability shapes behavior in those environments. We also complement the literature showing that cognitive ability predicts higher level-$k$ reasoning that operates at the level of a single round (see \citealp{BraGarGon_12}, \citealp{GilPro_16}, \citealp{AloBuc_21}, \citealp{Jin_21}, \citealp{FeGilPro_22}, and \citealp{BalRodVor_24}; although \citealp{GeoHeaWeb_15}, find mostly null results).

Finally, we provide evidence that, when level-$k$ reasoning operates at the level of a supergame strategy, subjects shift toward higher level-$k$ reasoning as they gain experience. Specifically, subjects are more likely to be classified as level-2, and less likely as level-1, in the final supergame compared to the first. This finding complements the literature that documents shifts toward higher level reasoning with experience when level-$k$ reasoning operates at the level of a single round (e.g., \citealp{Sta_96}, \citealp{DufNag_97}, \citealp{DanFefKub_12}, \citealp{GilPro_16}, \citealp{FerPitGar_25}).

The paper proceeds as follows. Section \ref{Sect:Design} describes the experimental design. Section \ref{Sect:Cluster} motivates the analysis by using a clustering algorithm to group subjects according to the similarity of their beliefs. Section \ref{Sect:Model} develops our model of level-$k$ reasoning in repeated games. Section \ref{Sect:Estimation} classifies subjects into levels of reasoning using the new notion of balance-optimal error tolerance. Section \ref{Sect:Regressions} uses regression analysis to study the effects of cognitive ability, personality, and experience on level-$k$ reasoning. Section \ref{Sect:Robustness} widens the type classification and investigates how level-$k$ types map empirically to beliefs and strategy choices. Section \ref{Sect:RobustMore} shows that our results are robust to the choice of error tolerance. Section \ref{Sect:Conclusion} concludes, while the Appendix provides further details and experimental screenshots.

\newpage

\section{Experimental design}\label{Sect:Design}

The data come from our earlier study, \cite{GilRos_24}. Here, we outline the experimental design (for further details and screenshots, see Section 2 and Appendices I and III of \citealp{GilRos_24}).

We ran 27 sessions with 394 student subjects at Purdue University's Vernon Smith Experimental Economics Laboratory. Subjects played 25 indefinitely repeated Prisoner's Dilemma supergames with random rematching. The stage-game payoff matrix (Figure \ref{Fig:Payoff}) follows \cite{DalFre_11}, with the return to joint cooperation $R\in\{32,40,48\}$, varied between subjects, and continuation probability $\delta=0.75$.

\begin{figure}[ht]
	\centering
	\begin{tikzpicture}[thick,scale=1.0, every node/.style={scale=1.0}]
    \draw (0,0) rectangle (4,4);
    \draw (2,0) -- (2,4);
    \draw (0,2) -- (4,2);
    \fontsize{12pt}{0pt}\selectfont
    \node [color=red] at (.75,1) {$50$ \ \ };
     \node [color=black] at (0.92,0.82) {\text{ , }};
    \node [color=blue] at (1.25,1) {$12$};
    \node [color=red] at (.72,3) {$R$ \ \ };
         \node [color=black] at (0.92,2.83) {\text{ , }};
    \node [color=blue] at (1.28,3) {$R$};
    \node [color=blue] at (3.25,3) {$50$};
        \node [color=black] at (2.93,2.82) {\text{ , }};
    \node [color=red] at (2.75,3) {$12$ \ \ };
    \node [color=red] at (2.75,1) {$25$ \ \ };
    \node [color=black] at (2.93,0.82) {\text{ , }};
    \node [color=blue] at (3.25,1) {$25$};
   \node [anchor=south, color=blue, align=center] at (1,4.2) {Cooperate};
    \node [anchor=south, color=blue, align=center] at (3,4.28) {Defect};
    \node [anchor=east, color=red, align=center] at (-0.21,3) {Cooperate};
    \node [anchor=east, color=red, align=center] at (-0.5,1) {Defect};
        \end{tikzpicture}
	\begin{minipage}{0.9\textwidth}
\vspace{0.5cm}
\footnotesize
\noindent
Notes: The experiment used labels `A' and `B' to represent Cooperate and Defect.
\end{minipage}
	\vspace{0.05cm}
		\caption{Stage-game payoff matrix}
	\label{Fig:Payoff}
		\end{figure}

\begin{figure}[ht]
\centering
\includegraphics[trim=0mm 0mm 0mm 0mm, width=0.75\textwidth]{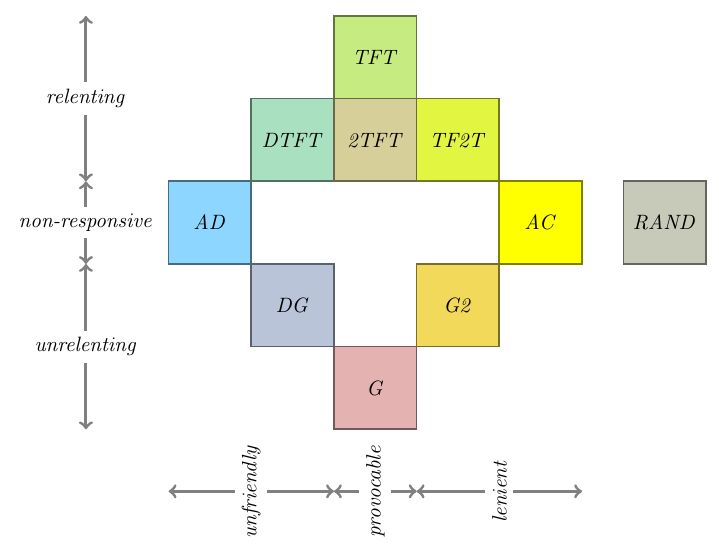}
\begin{minipage}{0.919\textwidth}
\vspace{0.5cm}
\footnotesize
\noindent
Notes:  AD: Always Defect; AC: Always Cooperate; G: Grim; G2: Lenient Grim 2; DG: Defect 1st round, then play G (ignoring own 1st-round defection); TFT: Tit-for-Tat; 2TFT: 2-Tits-for-1-Tat; TF2T: Tit-for-2-Tats; DTFT: Defect 1st round, then play TFT; RAND: Cooperate / Defect with probability $\frac{1}{2}$ in every round. \ref{Screen:Match}, page \pageref{Screen:Match}, provides the full wording of each strategy.
\end{minipage}
\vspace{0.1cm}
\caption{Description of the 10 strategies}
\label{Fig:Strategies}
\end{figure}

We elicited supergame strategies. At the start of each supergame, each subject chose one of 10 strategies to play the supergame, as described in Figure \ref{Fig:Strategies}. These strategies include the 8 most common identified via structural estimation using round-by-round choices in the IRPD (\citealp[Section I.D]{GilRos_24}), together with DG (the unrelenting analog of DTFT) and RAND (which randomizes each round). The subject’s chosen strategy and that of her opponent were then played out round-by-round on the subject’s screen. The supergame history (choices and payoffs) was displayed during the supergame and after the supergame ended (until the subject chose to continue to the next supergame). To preserve external validity, subjects were not told the strategy chosen by their opponent. \ref{Screen:Match}, page \pageref{Screen:Match}, provides an illustrative screenshot.

In Figure \ref{Fig:Strategies}, the horizontal axis categorizes the strategies according to when they first defect: `unfriendly’ strategies defect in the 1st round, while `provocable’ (`lenient’) strategies defect (do not defect) immediately in response to the opponent’s first defection. The vertical axis categorizes the strategies according to whether, after punishing a rival’s defection, they eventually relent and cooperate if the opponent cooperates.

\vspace{1cm}

To help subjects understand the 10 repeated-game strategies available in each of the 25 indefinitely repeated Prisoner's Dilemma supergames, subjects completed two forms of training (without incentives) before we elicited strategies in the 1st supergame. First, each subject played 10 practice supergames against herself using the direct-response method (choosing an action for herself and for the `other' in each round of each practice supergame). Second, subjects spent 3 minutes reading descriptions of the 10 available strategies, and then 5 minutes testing pairs of strategies against each other in training supergames played out round-by-round on the subject's screen (on average, testing 16.2 pairs, including repeats). Figures A.3 to A.6 in \cite{GilRos_24} provide illustrative screenshots.

We also elicited beliefs about supergame strategies. We elicited beliefs twice, in the 1st supergame and the 25th (final) supergame. Specifically, we elicited beliefs about the distribution of the 10 strategies, incentivized using the Quadratic Scoring Rule (QSR).\footnote{We gave subjects no feedback about the accuracy of their guesses. \citet[Appendix III.7]{GilRos_24} discuss properties of the QSR and evidence about effects of eliciting beliefs.} To prevent contamination of initial strategy choice, in the 1st supergame we elicited beliefs unexpectedly after subjects had chosen their strategy (but before the supergame was played out). \ref{Screen:Match}, pages \pageref{Screen:Belief1}-\pageref{Screen:Belief2}, provides illustrative screenshots.

We measured cognitive ability using matrix reasoning.\footnote{Matrix reasoning has been used to study how cognitive ability affects strategic behavior in a variety of games (e.g., \citealp{GilPro_16}, \citealp{ProRusSof_19,ProRusSof_22}, \citealp{FeGilPro_22}, \citealp{HerSch_22}).} At the beginning of the experiment, subjects completed the 11-item test of matrix reasoning from the International Cognitive Ability Resource (\citealp{DwoRevDoe_21}; https://icar-project.com), which is similar to Raven’s Progressive Matrices (\citealp{RavRavCou_00}), and which has been used in economics by, e.g., \cite{GilProKne_25}.  The test requires subjects to identify missing elements that complete visual patterns: Figure \ref{Fig:Matrix} in \ref{Sect:Further} provides a sample item. Matrix reasoning captures fluid intelligence (i.e., logical or analytical intelligence), and is a leading measure of cognitive ability that correlates strongly with general intelligence $g$ (\citealp{GraTho_04}, Box 1) and with performance on other complex cognitive tasks \citep{CarJusShe_90}. We standardize the matrix reasoning test scores, giving a cognitive ability measure with zero mean and unit standard deviation.

We measured personality using questions from the psychometric literature. At the beginning of the experiment, subjects completed a 52-item personality questionnaire. Based on the answers, and using principal factor analysis, we constructed five personality factors that measure:  anxiety; cautiousness; kindness; manipulativeness; and trust (for details see \citealp[Appendix III.9]{GilRos_24}). By construction, the personality factors are standardized (zero mean and unit standard deviation) and uncorrelated with each other. Furthermore, the personality factors all have correlations with cognitive ability that are smaller than 0.1.

Finally, we measured demographics at the end of the experiment. Specifically, we asked about age, gender, major, and whether the subject went to high school in the United States.

\newpage

\section{Belief clustering}\label{Sect:Cluster}

Before developing our model of level-$k$ reasoning in repeated games, we motivate our analysis by examining subjects' beliefs over strategies. To do so, we use a
clustering algorithm to group subjects into clusters according to the similarity of their beliefs.

Specifically, we use affinity propagation clustering  \citep{ FreDue_07}, which is a popular unsupervised machine learning clustering algorithm with two key advantages. First, affinity propagation determines the optimal number of clusters endogenously. Second, within each cluster, affinity propagation selects one subject’s beliefs to be the ‘exemplar’ beliefs that are representative of beliefs in that cluster. Affinity propagation has been used previously by, e.g., \cite{RosRom_18} to cluster repeated game strategies.

Panel (a) of Table \ref{Tab:Beliefs} reports the output of the affinity propagation clustering algorithm for the beliefs over strategies elicited in the 1st supergame. Cluster 1, the largest cluster with 56 subjects, has exemplar beliefs that are exactly uniform. Clusters 2, 3, 5, 6 and 7 also have close-to-uniform beliefs (specifically, these clusters have exemplar beliefs whose mean-squared deviation from uniform does not exceed 25).

Together, the 187 subjects in these 6 clusters with close-to-uniform beliefs in the 1st supergame help to motivate the development of our model of level-$k$ reasoning in repeated games, since uniform beliefs are consistent with a level-1 type who believes others are strategically unsophisticated (level-0) and therefore randomize uniformly over strategies. Our level-$k$ analysis is further motivated by the finding that our subjects generally choose strategies that perform well given their beliefs: in the 1st supergame, around half of the subjects choose a strategy that achieves at least 95\% of the expected payoff from the best response to their beliefs, while around a quarter of the subjects choose a strategy that is a perfect best response (panel (a) of Figure \ref{Fig:CDF} in \ref{Sect:Further} provides the cumulative distribution function).

\begin{table}[ht]
\centering
\begin{subtable}{1.07\textwidth}
    \centering
    \includegraphics[trim=12mm 111mm 2mm 6.5mm, clip, width=0.95\textwidth]{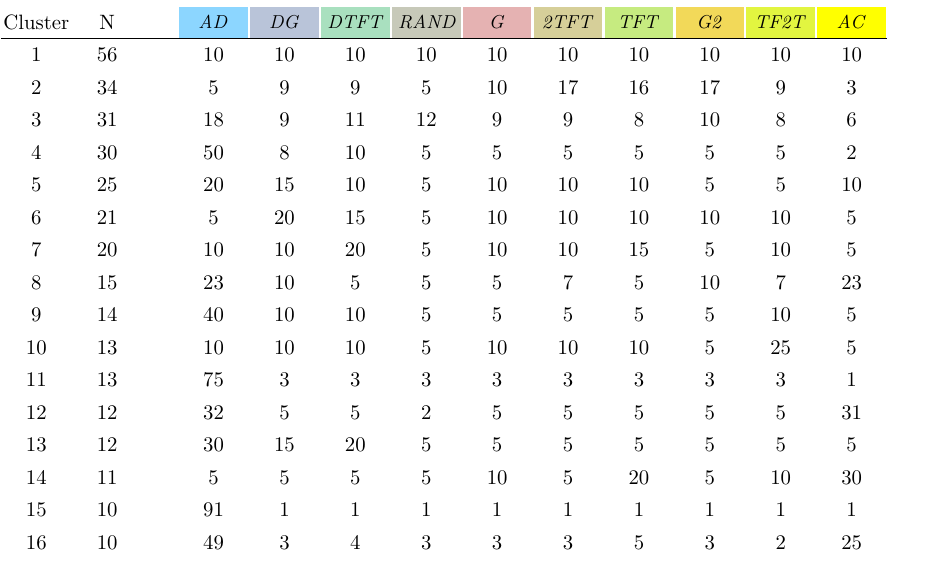}
       \vspace{-0.0cm}
    \caption{Supergame 1}
    \label{Tab:BeliefsS1}
    \vspace{0.7cm}
\end{subtable}
\begin{subtable}{1.07\textwidth}
    \centering
    \includegraphics[trim=12mm 102mm 2mm 6.5mm, clip, width=0.95\textwidth]{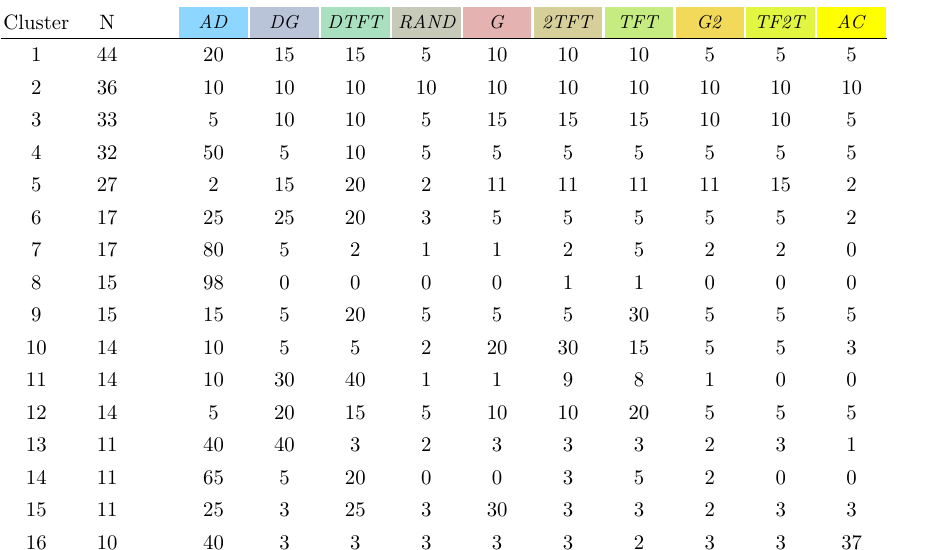}
    \vspace{0.25cm}
    \caption{Supergame 25}
    \label{Tab:BeliefsS25}
      \vspace{0.8cm}
\end{subtable}
\begin{minipage}{0.95\textwidth}
\footnotesize
\noindent
Notes: For each cluster, the table reports the exemplar beliefs. We report here clusters with $N \geq 10$; Tables \ref{Tab:BeliefsS1_All} and \ref{Tab:BeliefsS25_All} in \ref{Sect:Further} report all the clusters (none of the smaller clusters has exemplar beliefs with mean-squared deviation from uniform $\leq 25$). We implemented affinity propagation clustering in R using the APCluster package \citep{BodKotHoc_11}, and using default values for all arguments. \end{minipage}
      \vspace{0.2cm}
\caption{Belief clusters}
\label{Tab:Beliefs}
\end{table}

\clearpage

Panel (a) of Table \ref{Tab:Beliefs} also shows other interesting patterns of beliefs. For example, a number of clusters have exemplar beliefs that place 50\% or more weight on AD (e.g., clusters 4, 11 and 15), while others have exemplars that place 50\% or more weight on AC (clusters 17, 22 and 23; see Table \ref{Tab:BeliefsS1_All} in \ref{Sect:Further}). An interesting belief pattern also emerges in which AD and AC both attract relatively high probability weights, with close-to-uniform beliefs among the remaining strategies (e.g., clusters 8, 12 and 16): these clusters suggest that some subjects’ initial beliefs tend to gravitate toward simple memory-0 strategies.

Recall that we also elicited beliefs over strategies in the 25th (final) supergame: panel (b) of Table \ref{Tab:Beliefs} reports the corresponding belief clusters. Clusters 1, 2 and 3 have close-to-uniform beliefs (again, these clusters have exemplar beliefs whose mean-squared deviation from uniform does not exceed 25), and we find 113 subjects in these clusters. The decline in the number of subjects in clusters with close-to-uniform beliefs, from 187 in the 1st supergame to 113 in the 25th supergame, suggests that some subjects moved from level-1 to higher level reasoning, which in turn motivates our later analysis of how experience affects level-$k$ reasoning in Section \ref{Sect:Regressions}.\\

\clearpage

\section{Level-$k$ model: Description}\label{Sect:Model}

In this section, we develop a model of level-$k$ reasoning in repeated games that incorporates choices over supergame strategies as well as beliefs about the supergame strategies chosen by others. We build on \cite{Nag_95}'s level-$k$ model of boundedly rational reasoning, in which the level-0 type is strategically unsophisticated, the level-1 type best responds to the belief that all other players are level-0, the level-2 type best responds to the belief that all other players are level-1, and so on. As described by \cite{CraCosIri_13}'s review of the literature, level-$k$ rules ``anchor beliefs in an instinctive reaction to the game and then adjust them via a small number of iterated best responses.''\footnote{As further described by \cite{Nag_95}, the level-1 type ``forms first-order beliefs on the behavior of the other players. He thinks that others select a number at random, and he chooses his best response to this belief,'' while the level-2 type ``forms second-order beliefs on the first-order beliefs of the others...'' }

The model describes behavior in one repeated game (i.e., in one supergame). We assume that players choose from a finite set of supergame strategies. We make this assumption for simplicity and to match our experimental data (described in Section \ref{Sect:Design}), noting also that evidence from strategy estimation based on round-by-round choices in the indefinitely repeated Prisoner's Dilemma shows that most choices are consistent with a limited set of strategies \citep{DalFre_18}. We further assume that players form a belief about the supergame strategies chosen by others, in the form of a probability distribution over the finite set of strategies.

\setcounter{assumption}{-1}
\begin{assumption}
{The level-0 type randomizes uniformly over the finite set of strategies.}
\end{assumption}

Uniform randomization by the strategically unsophisticated level-0 type is a common assumption in the level-$k$ literature (e.g., \citealp{CosCraBro_01}, \citealp{CraIri_07}, \citealp{FudLia_19}). \cite{CraCosIri_13}'s review notes that: ``In most applications ... \textit{L0} is assumed to be uniform random over others' feasible decisions...''

\begin{assumption}\label{Ass:1}
The level-1 type holds beliefs within distance $\eta$ of uniform and chooses a strategy within distance $\eta$ of a best response to her beliefs, where $\eta \in [0,1]$ indexes the degree of error tolerance.
\end{assumption}

As is standard in the level-$k$ literature, we allow types to make errors (e.g., \citealp{HoCamWei_98}, \citealp{GilPro_16}, and \citealp{HoSu_21}, allow noise in choices, while \citealp{DanFefKub_12}, allow noise in beliefs). \cite{CraCosIri_13}'s review notes that: ``In empirical applications, it is assumed that \textit{L1} and higher types make errors...'' In particular, $\eta \in [0,1]$ indexes the degree of error tolerance in terms of distance, for both beliefs and choices. When $\eta=0$, the error-free level-1 type believes that all other players are level-0, and so she holds uniform beliefs while choosing a best response to her uniform beliefs. When $\eta>0$, the level-1 type's beliefs can diverge from uniform by at most a distance of $\eta$, and the level-1 type's strategy can diverge from a best response to her beliefs by, again, at most a distance of $\eta$.\footnote{Noise in beliefs could stem from subjects whose beliefs deviate in a systematic way from the exact level-$k$ belief prediction or whose beliefs deviate randomly every time they play the game. Either way, in our level-$k$ model, players best respond (with noise) to their noisy belief. This is different from \cite{CosWei_08}, who assume that a latent belief drives behavior and is reported with noise. In our setting with learning and only two observations per subject, we cannot recover latent beliefs.}

We use payoff space to measure distance. This choice aligns naturally with our experimental data, where both choices and beliefs were incentivized, but in other settings alternative distance metrics could be more appropriate. Specifically, we say that a player's strategy $s$ is distance $d_s \in [0,1]$ from the best response to her beliefs if $s$ achieves expected payoffs, given her beliefs, that are a proportion $1-d_s$ of the expected payoffs from the best response to her beliefs.\footnote{With multiple best responses, this payoff-based distance is the same across all best responses.} Furthermore, we say that a player's beliefs $b$ are distance $d_b \in [0,1]$ from distribution $x$ if reporting $b$ when the true distribution is $x$ achieves expected payoffs that are a proportion $1-d_b$ of the expected payoffs from reporting $x$ (as noted in Section \ref{Sect:Design}, beliefs were incentivized using the QSR). For comparability across the domains of choices and beliefs when using payoff space to measure distance, we normalize payoffs to range from 0 to 1 within each domain.\footnote{Alternatively, the model could be generalized to allow separate error tolerance parameters, one for choices and another for beliefs.}

\begin{assumption}\label{Ass:2}
The level-2 type holds beliefs that are within distance $\eta$ of a distribution that places probability weight only on strategies that are consistent with level-1 behavior, and chooses a strategy within distance $\eta$ of a best response to her beliefs.
\end{assumption}

Just like for the level-1 type, $\eta \in [0,1]$ indexes the degree of error tolerance in terms of distance, for the level-2 type's beliefs and choices. We define ``strategies that are consistent with level-1 behavior'' as strategies that are within distance $\eta$ of a best response to uniform beliefs. This definition implies that the level-2 type allows for potential error in the level-1 type's choices. Previous models in the level-$k$ literature that include higher level types who take into account noisy behavior by lower level types include \cite{StaWil_94}, \cite{HoCamWei_98}, and \cite{GoeLouZha_18}.\footnote{As described by \cite{CraCosIri_13}'s review, the level-2 type in \cite{StaWil_94}'s model is ``best responding to a noisy \textit{L1}.'' In \cite{HoCamWei_98}'s model: ``Level-$L$ players are assumed to believe that all other players (besides themselves) choose from the Level $L-1$ density,'' where the lower level density incorporates noisy choice by the lower level agents. \cite{GoeLouZha_18} use a level-$k$ model with noisy introspection where ``level-1 makes a noisy best response to level-0, level-2 makes a noisy best response to the noisy play of level-1, etc.''} The definition further implies that the level-2 type assumes that the level-1 type holds uniform beliefs: thus, the level-2 type does not allow for error in the level-1 type's beliefs. We make this assumption for conceptual and computational simplicity, noting that \cite{DanFefKub_12} also assume a level-$k$ type with noisy beliefs who does not allow for noise in the beliefs of the level-$(k-1)$ type.

Although we do not include types higher than level-2 in our estimation (see Section \ref{Sect:Estimation} for discussion), the model could be extended to include higher types, with a \mbox{level-$k$} type (for $k>2$) holding beliefs that are within distance $\eta$ of a distribution that places probability weight only on strategies that are compatible with level-($k-1)$ behavior, while also choosing a strategy within distance $\eta$ of a best response to her beliefs.

\newpage

\section{Level-$k$ model: Estimation}\label{Sect:Estimation}

\subsection{Estimation methodology}

We follow, for example,  \cite{NyaSch_02}, \cite{BelKroSoe_08} and \cite{Spi_12} by using elicited beliefs, together with choices, to estimate a structural model of strategic behavior. We estimate our level-$k$ model using the data from \cite{GilRos_24} described in Section \ref{Sect:Design}. Recall that we elicited beliefs about supergame strategies only in the 1st supergame and 25th (final) supergame, and so we estimate the model twice, first using the data from the 1st supergame, and second using the data from the 25th supergame.

We include only the level-1 and level-2 types in our estimation. We exclude higher types because in our setting with choices and beliefs over 10 strategies, and where subjects are given only the stage-game payoff matrix, the computational burden on higher types makes higher level behavior somewhat implausible. Furthermore, in mostly simpler settings, the level-$k$ literature finds most weight on level-1 and level-2 behavior.\footnote{\cite{CraIri_07} summarize the level-$k$ literature as follows: ``The estimated distribution tends to be stable across games, with most of the weight on \textit{L1} and \textit{L2}.'' For example, using first-round choices in the Beauty Contest game with $p<1$, \cite{DufNag_97} find that level-1 and level-2 types make up 73\% of the population, with only a very small proportion ($<5\%$) of level-3 types.} We exclude the level-0 type because we cannot use beliefs to identify level-$0$ behavior. Furthermore, much of the level-$k$ literature assumes that the level-0 type has zero frequency in the population, serving only to anchor the thinking of higher level types (e.g., \citealp{CosCra_06}, \citealp{CraIri_07b}, \citealp{HoSu_13}). \cite{CraCosIri_13}'s review notes that ``there is no presumption that \textit{L0} players exist: \textit{L0} is simply \textit{L1}'s model of others...'' and that ``the estimated frequency of \textit{L0} is usually zero or small.''

We follow a common approach in the level-$k$ literature by estimating the model subject by subject (e.g., \citealp{Nag_95}, \citealp{CosCra_06}, \citealp{AloBuc_21}). Recall that $\eta \in [0,1]$ indexes error tolerance, for both choices and beliefs, with error tolerance increasing in $\eta$. We begin by estimating our level-$k$ model for every value of $\eta$ on a fine discrete grid.\footnote{We use the grid $\eta \in \{0,0.00001, 0.00002, ... , 1\}$. Payoff calculations are based on analytical calculations of expected payoffs for every pair of strategies, but are subject to computer precision. To establish whether a subject's beliefs are consistent with the level-2 type, we use numerical optimization to minimize distance between the subject's beliefs and distribution $z$, subject to the constraint that $z$ places probability weight only on strategies that are consistent with level-1 behavior (specifically, we use the SLSQP constrained optimization algorithm from the scipy.optimize Python library, with multiple random initializations).} Specifically, for either Supergame 1 or Supergame 25, and for every value of $\eta$, we classify each subject into one of four mutually exclusive categories, as follows (Assumptions \ref{Ass:1} and \ref{Ass:2} in Section \ref{Sect:Model} describe the level-1 and level-2 types).
\vspace{0.5cm}

\noindent \textit{Level-$k$ classification:}

\begin{itemize}[leftmargin=4.6mm]

\item \textbf{L1}: Choices and beliefs jointly consistent only with the level-1 type.

\item \textbf{L2}: Choices and beliefs jointly consistent only with the level-2 type.

\item \textbf{Both}: Choices and beliefs jointly consistent with both the level-1 and level-2 types.

\item \textbf{Neither}: Choices and beliefs jointly consistent with neither the level-1 nor level-2 type.

\end{itemize}

\vspace{0.3cm}

Note that we use the term `estimation' loosely: for each value of error tolerance $\eta$, we simply classify each subject into a category. This approach allows us to classify subjects for any level of error tolerance, which in turn motivates our notion of balance-optimal error tolerance, as described below. We follow, e.g., \cite{Nag_95} in classifying subjects based on distance, while our approach contrasts with, e.g., \cite{GilPro_16} who estimate a mixture-of-types level-$k$ model of learning using maximum likelihood (Section \ref{Sect:ComparisonML} compares our classification methodology to maximum likelihood approaches).

\subsection{Results}

\begin{figure}[ht]

    \centering

    \begin{subfigure}{0.9\textwidth}
        \centering
        \includegraphics[width=\linewidth]{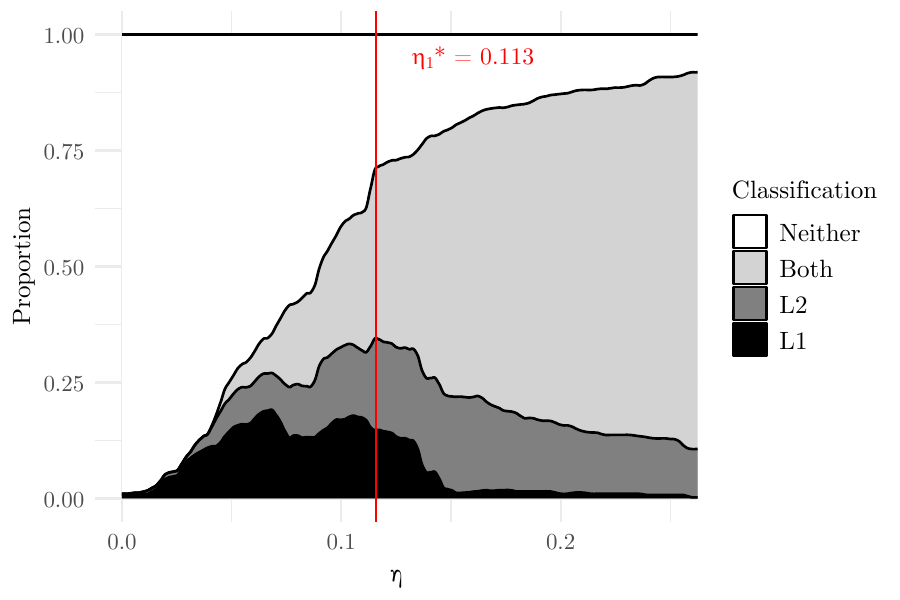}
        \caption{Supergame 1}
        \label{Fig:Stacked1}
    \end{subfigure}

    \vspace{0.8cm}

    \begin{subfigure}{0.9\textwidth}
        \centering
        \includegraphics[width=\linewidth]{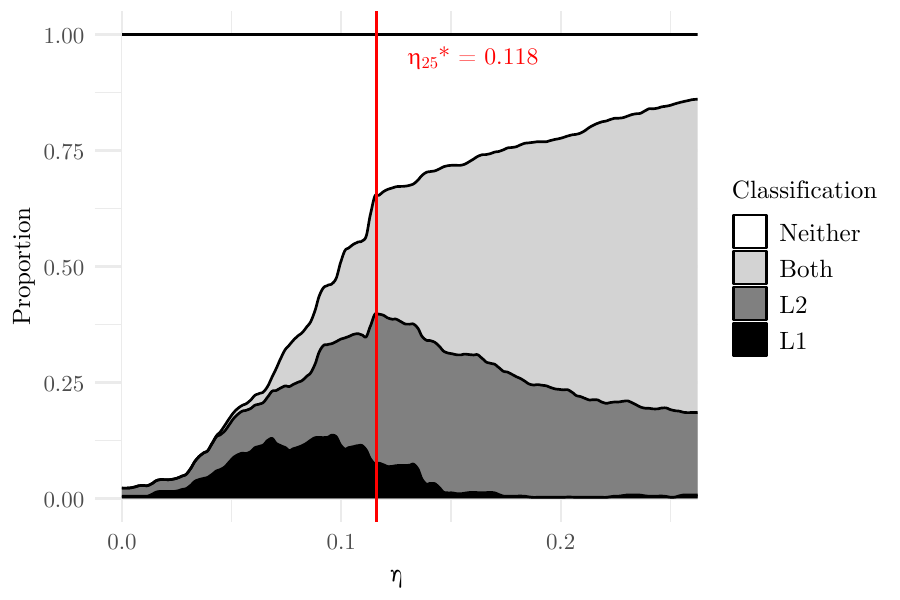} %
        \caption{Supergame 25}
        \label{Fig:Stacked25}
    \end{subfigure}

    \vspace{0.8cm}

    \begin{minipage}{0.85\textwidth}
    \footnotesize
    \noindent
    Notes: The plots classify all 394 subjects and are smoothed using LOESS with a span of 0.01.
    \end{minipage}

      \vspace{0.3cm}

    \caption{Level-$k$ classification for range of values of error tolerance $\eta$}
    \label{Fig:Stacked}
\end{figure}

The stacked graphs in Figure \ref{Fig:Stacked} present the classification results for the 1st supergame (panel (a)) and the 25th supergame (panel (b)), for values of error tolerance $\eta \leq 0.25$. When error tolerance is near zero, we classify few subjects as L1 or L2: unsurprisingly, few subjects conform to the level-$k$ model when we allow near zero noise in choices or beliefs. As error tolerance increases from zero, at first we classify more and more subjects as L1 or L2. However, once error tolerance becomes large enough, the proportion of subjects classified as L1 or L2 begins to fall as $\eta$ increases further. The reason is that when we tolerate too much error, the level-$k$ model loses predictive power: a wide range of behavior becomes consistent with both the level-1 type and the level-2 type, pushing many subjects to be classified as Both. This discussion motivates the notion of balance-optimal error tolerance.

\begin{definition}
The balance-optimal error tolerance $\eta^*$ is the value of $\eta$ that maximizes the combined share of subjects classified as L1 or L2.
\end{definition}

That is, the balance-optimal error tolerance maximizes the share of subjects classified uniquely as level-1 or level-2. The balance-optimal error tolerance strikes a balance: tolerating enough error for the level-$k$ model to explain observed behavior, while not tolerating so much error that the behavior of level-1 and level-2 types overlaps excessively. Next, we report the level-$k$ classification at the balance-optimal error tolerance.

\begin{result}\label{R1}
For the 1st supergame, the balance-optimal error tolerance $\eta^*_1=0.113$, and at $\eta^*_1$ we classify 14.7\% of subjects as L1, 20.1\% as L2, and 34.5\% as Both.
\end{result}

\begin{result}\label{R2}
For the 25th supergame, the balance-optimal error tolerance $\eta^*_{25}=0.118$, and at $\eta^*_{25}$ we classify 7.6\% of subjects as L1, 32.7\% as L2, and 25.6\% as Both.
\end{result}

\vspace{0.15cm}

Comparing Result \ref{R1} with Result \ref{R2}, and also comparing panels (a) and (b) of Figure \ref{Fig:Stacked} across the range of values of error tolerance, we can see that the level of reasoning increases with experience: as we move from the 1st supergame to the 25th (final) supergame, the proportion of subjects classified as L1 goes down, while the proportion classified as L2 goes up. In Section \ref{Sect:Regressions}, we conduct econometric tests to confirm that this increase in level of reasoning with experience is statistically significant.

At the balance-optimal error tolerance we classify many subjects as Both. In Section \ref{Sect:Robustness}, we introduce an alternative, wider level-$k$ classification that reassigns these subjects to L1 or L2, using the fact that each of these subjects' beliefs are consistent with only the level-1 type or only the level-2 type when error tolerance $\eta$ is lowered sufficiently. Section \ref{Sect:Robustness} shows that our findings remain robust under this wider classification.

Finally, in Section \ref{Sect:RobustMore} we show that our results are robust to the choice of error tolerance. First, to account for sampling variability, we bootstrap the balance-optimal error tolerance and re-run our analysis using error tolerances toward the ends of the bootstrapped range. Second, we classify all subjects as L1 or L2 without using the balance-optimal error tolerance at all.

\clearpage

\subsection{Importance of beliefs}

We emphasize that our level-$k$ classification relies on using subjects’ elicited beliefs, for two reasons. First, since beliefs consistent with different level-$k$ types can rationalize the same strategy choice, we use beliefs to distinguish level-$k$ types. For example, consider subject \textit{q2hmdlzt} in the $R=32$ treatment who chooses AD. We classify this subject as L2 because AD is a best response to the subject's level-2 beliefs that place 100\% weight on AD. Without beliefs, we could not distinguish this subject from a level-1 type whose best response to uniform beliefs is also AD. We note that this difficulty in distinguishing level-$k$ types without using elicited beliefs is particularly problematic in our complex setting. Given this complexity, we tolerate noise in beliefs (through the error tolerance parameter $\eta$), but this noise expands the set of strategies that can be rationalized by different level-$k$ beliefs. Second, since the same strategy choice could arise from beliefs consistent with level-$k$ reasoning or from some other process, we use beliefs to identify whether subjects are level-$k$ types or not. For example, consider subject \textit{yo91q5ok} in the $R=32$ treatment who chooses AD. We classify this subject as Neither because her elicited beliefs are far from either level-1 or level-2 beliefs, placing 50\% weight on AD and 50\% weight on 2TFT. Without beliefs, and based only on choosing AD, we could not distinguish this subject from a level-$k$ type.

\subsection{Further discussion of our methodology and connection to maximum likelihood}\label{Sect:ComparisonML}

Our classification methodology has three advantages. First, it is direct and transparent. Fixing the error tolerance $\eta$, each subject's classification follows directly from her observed choices and beliefs, and we can observe clearly how the classification varies with the amount of error that we tolerate (Figures \ref{Fig:Stacked1} and \ref{Fig:Stacked25}). Second, error tolerance has a direct economic interpretation as the maximal proportional payoff loss allowed (for choices and for beliefs). Third, by separating level-$k$ from non-level-$k$ behavior in a structured data-driven way, our notion of balance-optimal error tolerance guards against classifying as level-1 or level-2 those subjects whose behavior in fact arises from some other process. Overall, our approach requires no parametric assumptions on the distribution of errors, no probabilistic allocation of subjects to types, and no arbitrary specification of non-level-$k$ types.

By contrast, mixture-of-types maximum likelihood estimation is more complex, requiring the modeling of joint choice and belief distributions with assumptions on the error structure. Furthermore, allocating subjects to types requires using noisy posterior probabilities. Importantly, distinguishing level-$k$ from non-level-$k$ behavior within the maximum likelihood framework requires specifying particular non-level-$k$ types. For example, \cite{CosCraBro_01} include dominance, equilibrium, and sophisticated types. In our approach, the Neither category instead accommodates non-level-$k$ behavior without having to take a stand on what that behavior is.

We classify based on distance measured in payoff space. This contrasts with approaches that classify based on distance measured in the space of actions, such as the distance between chosen and level-$k$ predicted guesses in the Beauty Contest (e.g., \citealp{Nag_95}, \citealp{AgrPotSch_12}, \citealp{AloBuc_21}). Since we measure distance in payoff space, our classification methodology has a close connection to maximum likelihood that directly classifies each individual (e.g., \citealp{CosCra_06}), rather than estimating a mixture model. Noting that each subject's choice is a fixed distance from the best response to her beliefs, and under the additional assumption that errors in choices and beliefs are independent, the maximum likelihood comparison of the level-1 and level-2 types reduces to comparing the distance in payoff space from the subject's beliefs to each type's level-$k$ predicted beliefs (this holds true for any logit precision parameter). Thus, at any error tolerance $\eta$ used to calculate the level-$k$ predicted beliefs, and taking the subjects we classify as L1 or L2, maximum likelihood yields exactly the same classification.

\newpage

\section{Effects of cognitive ability and experience}\label{Sect:Regressions}

In this section, we turn to regression analysis to study the effects of cognitive ability, personality, and experience on level-$k$ reasoning. The regressions reported in Table \ref{Tab:Regressions} use the level-$k$ classification from Section \ref{Sect:Estimation} reported in Results \ref{R1} and \ref{R2}, and are based on data from the 1st and 25th supergames (the table notes provide full details).

In the 1st column, the dependent variable is an indicator for being classified as L2, and the sample includes only observations in which a subject is classified as either L1 or L2. Thus, the regression measures the effects of cognitive ability and personality on the probability of being classified as L2, conditional on being L1 or L2. The results show that a 1 standard deviation increase in cognitive ability is associated with a 9 percentage point increase in the probability of being classified as L2 ($p=0.002$).

In the 2nd column, we further measure the effect of experience. The regression is the same as in the 1st column, except that we add an indicator for the 25th (final) supergame. The coefficient on this indicator shows that the probability of being classified as L2 in the 25th supergame is 22 percentage points higher than in the 1st supergame ($p < 0.001$).

In the columns headed ``All Data'', we expand the sample to include all observations (that is, the sample now further includes observations in which a subject is classified as Both or Neither). The 5th and 6th columns (All Data, L2) show that cognitive ability continues to predict the probability of being classified as L2 ($p=0.007$). In the 6th column, the coefficient on the 25th supergame shows that the probability of being classified as L2 continues to be higher with experience ($p = 0.001$). Finally, the 3rd and 4th columns (All Data, L1) show weaker, negative effects of cognitive ability and experience on the probability of being classified as L1.

We summarize the main findings from Table \ref{Tab:Regressions} as follows:

\begin{result}\label{R3}
Cognitive ability predicts higher level-$k$ reasoning.
\end{result}

\begin{result}\label{R4}
With experience, subjects shift toward higher level-$k$ reasoning.
\end{result}

\vspace{0.15cm}

\begin{table}[ht]
\centering
\includegraphics[trim=0mm 0mm 0mm 0mm, width=0.95\textwidth]{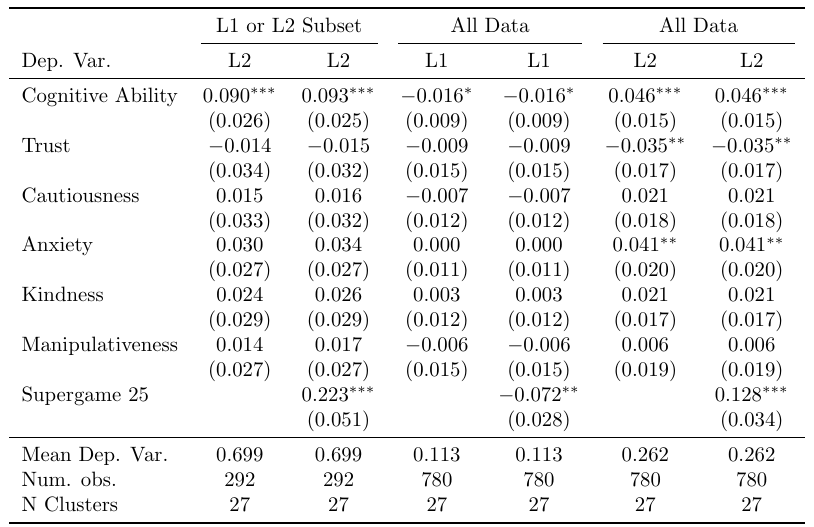}
\begin{minipage}{0.93\textwidth}
\vspace{0.3cm}
\footnotesize
\noindent
Notes: Col. 1 reports an OLS regression where the dependent variable is an indicator for being classified as L2. The regression uses data from Supergame 1 and Supergame 25, and the sample includes only observations in which a subject is classified as either L1 or L2 at the balance-optimal error tolerance for the given supergame (subjects may therefore appear zero, one, or two times). Col. 2 adds an indicator for Supergame 25. Cols. 5 and 6 are the same as Cols. 1 and 2, except that the sample includes all observations. Cols. 3 and 4 are the same as Cols. 5 and 6, except that the dependent variable is an indicator for being classified as L1. Cognitive ability and personality measures are standardized. All regressions include controls for treatment and demographic characteristics (see Section \ref{Sect:Design}), specifically binary indicators for $R=40$, $R=48$, and for each characteristic. As in \cite{GilRos_24}: (i) we binarize major into STEM / not STEM; (ii) we exclude from the regressions 4 subjects who answered ``prefer not to say'' to one or more demographic questions. Heteroskedasticity-robust standard errors clustered at the session level are in parentheses. $^{***}$, $^{**}$ and $^*$ denote significance at the 1\%, 5\% and 10\% levels (two-sided tests).
\end{minipage}
\vspace{0.3cm}
\caption{Effects of cognitive ability, personality, and experience on level-$k$}
\label{Tab:Regressions}
\end{table}

\clearpage

We find no evidence of an interaction between the effects of cognitive ability and experience. When we add an interaction between cognitive ability and the indicator for the 25th supergame to the regressions reported in the 2nd, 4th and 6th columns of Table \ref{Tab:Regressions}, in every case the coefficient on the interaction is far from statistical significance ($p>0.7$).

Finally, we find little systematic evidence that personality traits predict level-$k$ reasoning. We find some evidence that subjects higher in anxiety are more likely to be classified as L2: the effect of anxiety is significant at the 5\% level in the 5th and 6th columns, and we find stronger effects of anxiety when we widen the classification of L1 and L2 (see Table \ref{Tab:RegressionsWide} in Section \ref{Sect:Robustness}).\footnote{Anxiety is one facet of neuroticism. \cite{GilPro_16} find a positive association between a factor that loads on both agreeableness and emotional stability (the reverse of neuroticism) and higher level-$k$ in a repeated Beauty Contest game.}
Although trust is statistically significant at the 5\% level in the 5th and 6th columns, it is not significant in the other columns, and the effects of trust are no longer significant at the 5\% level once we widen the classification of L1 and L2.\footnote{Although not our focus of interest, our dataset includes gender as a demographic control: perhaps intriguingly, the coefficient on the gender control shows that male subjects are more likely to be classified as L2 in our data ($p<0.01$ in the regressions reported in the 5th and 6th columns of Table \ref{Tab:Regressions}). In one-shot games, \citet{HynTerVak_22} categorize subjects into `stayers' whose level-$k$ beliefs are stable across games and `movers' whose level-$k$ beliefs change across games, finding that stayers have level-1 beliefs and are more likely to be female, while movers tend to have higher level beliefs and are more likely to be male (the results on gender are in Appendix C.3). Again in one-shot games, and using the cognitive hierarchy model, \cite{CamHoCho_04} also find higher level thinking among males. \cite{BraGarGon_12}, however, find no gender difference in level-$k$.}\\

\clearpage

\section{Wider classification and behavior of L1 and L2}\label{Sect:Robustness}

In this section, we widen the classification of L1 and L2 to include subjects that we previously classified as Both, and we show that our earlier results are robust to this wider classification. We also use the wider classification to investigate how L1 and L2 map empirically to beliefs and choices.

Section \ref{Sect:Estimation} classified each subject into one of four categories: L1, L2, Both, or Neither. Results \ref{R1} and \ref{R2} report the level-$k$ classification: at the balance-optimal error tolerance for the given supergame, we classified many subjects as Both. Consider a subject classified as Both: this subject has choices and beliefs that are jointly consistent with both the level-1 and level-2 types. As we lower error tolerance $\eta$ below the balance-optimal level $\eta^*$, eventually we reach a value at which the subject's beliefs are consistent with one and only one of the level-1 and level-2 types (Assumptions \ref{Ass:1} and \ref{Ass:2} define belief consistency). Here, we reassign the subject from Both to L1 (L2) if her beliefs are consistent only with the level-1 (level-2) type at this lower $\eta$. Using this methodology, we are able to reassign all subjects previously in the Both category as L1 or L2.

Based on this wider classification, for the 1st supergame we now classify 41.6\% of subjects as L1 and 27.7\% as L2, and for the 25th supergame we now classify 28.9\% of subjects as L1 and 37.1\% as L2. By construction, these percentages sum to the total of L1, L2 and Both from Results \ref{R1} and \ref{R2}. Using the wider classification, we continue to find that the level of reasoning increases with experience: as we move from the 1st supergame to the 25th (final) supergame, the proportion of subjects classified as L1 continues to go down, while the proportion classified as L2 continues to go up.

\begin{table}[ht]
\vspace{0.4cm}
\centering
\includegraphics[trim=0mm 0mm 0mm 0mm, width=0.95\textwidth]{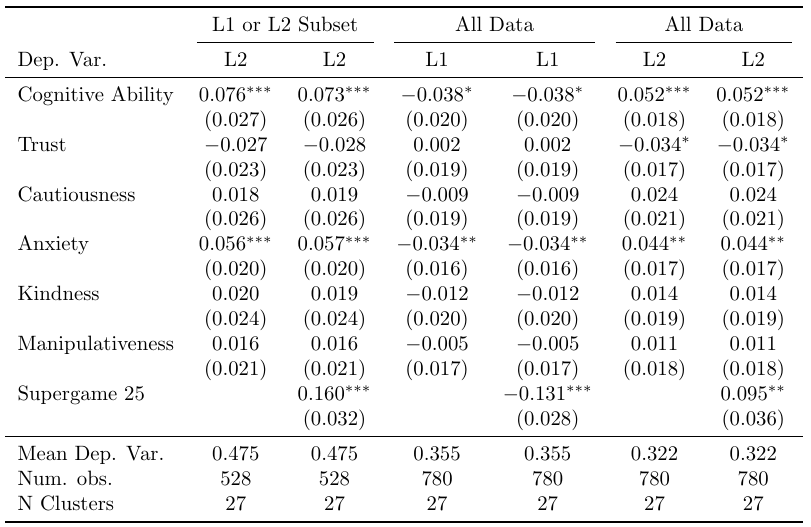}
\begin{minipage}{0.93\textwidth}
\vspace{0.3cm}
\footnotesize
\noindent
Notes: See the notes to Table \ref{Tab:Regressions}. Here, we run the same regressions, except that we use the wider classification of L1 and L2 described in the main text (which reassigns subjects in Both to L1 or L2). Heteroskedasticity-robust standard errors clustered at the session level are in parentheses. $^{***}$, $^{**}$ and $^*$ denote significance at the 1\%, 5\% and 10\% levels (two-sided tests).

\end{minipage}
\vspace{0.3cm}
\caption{Effects of cognitive ability, personality, and experience on level-$k$:\\ Wider classification of L1 and L2}
\label{Tab:RegressionsWide}
\end{table}

With the wider classification of L1 and L2, we continue to find statistical support for Results \ref{R3} and \ref{R4} from Section \ref{Sect:Regressions}.  Table \ref{Tab:RegressionsWide} reports the same regressions as in Table \ref{Tab:Regressions}, but now using the wider classification. In particular, the positive effects of cognitive ability and experience on the probability of being classified as L2 continue to be statistically significant, with $p<0.05$ in all cases. As previously noted in Section \ref{Sect:Regressions}, we also find that subjects higher in anxiety are more likely to be classified as L2.

\clearpage

Next, we use the wider classification to investigate how L1 and L2 map empirically to beliefs and choices, split by treatment. Specifically, Table \ref{Tab:MapS1} reports the empirical beliefs and strategy choices that we observe in the data for each treatment-category pair. We use the wider classification for this purpose because splitting by treatment with four categories (L1, L2, Both, and Neither) cuts the data too finely, with some treatment-category pairs containing few subjects. Table \ref{Tab:MapS1} reports data for the 1st supergame: Table \ref{Tab:MapS25} in \ref{Sect:Further} shows that the patterns described below persist in the 25th supergame.

Consistent with our level-$k$ model, Table \ref{Tab:MapS1} shows that the average belief weights of L1 subjects are approximately uniform.\footnote{Given our level-$k$ model and classification methodology, this pattern is not surprising.} When the return to joint cooperation is low ($R=32$), L1 subjects favor unfriendly strategies that defect in the 1st round (AD, DTFT and DG, which are the three highest-expected-payoff strategies given uniform beliefs when $R=32$). When the return to joint cooperation is high ($R=48$), L1 subjects shift toward strategies that cooperate in the 1st round, with G2 being the most popular (G2 is the unique best response given uniform beliefs when $R=48$).

When the return to joint cooperation is low ($R=32$), Table \ref{Tab:MapS1} shows that L2 subjects place most belief weight on unfriendly strategies, broadly matching L1 behavior (L1 choose unfriendly strategies 76\% of the time, while L2's beliefs place 72\% weight on unfriendly strategies, skewed toward AD). Consistent with this pessimism, L2 choose AD 84\% of the time. When the return to joint cooperation is high ($R=48$), L2 beliefs become substantially more optimistic, in line with L1’s more cooperative strategy choices. Consistent with this optimism, L2 subjects never choose unfriendly strategies.

Finally, Table \ref{Tab:MapS1} provides evidence that the Neither category includes some strategically unsophisticated subjects. First, the average beliefs of subjects classified as Neither place slightly \textit{more} weight on unfriendly strategies when the return to joint cooperation is high ($R=48$) than when it is low ($R=32$), moving in the opposite direction to incentives. Second, only subjects classified as Neither choose AC when $R=32$ or AD when $R=48$. Third, they choose RAND more frequently than L1 or L2 subjects.

\newpage

\begin{table}[ht]
\centering
\begin{subtable}{\textwidth}
    \centering
    \includegraphics[trim=0mm 0mm 0mm 0mm, clip, width=1.00\textwidth]{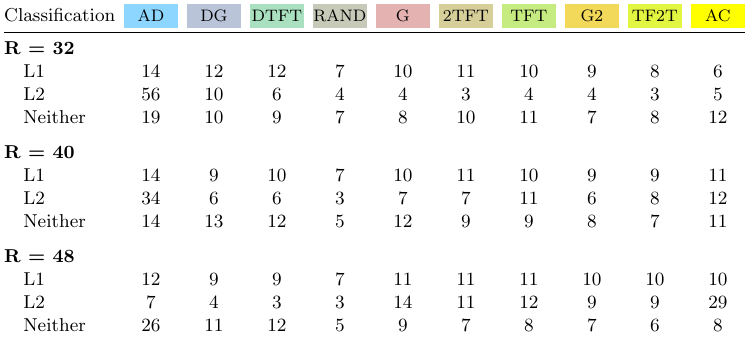}
       \vspace{-0.5cm}
    \caption{Beliefs in Supergame 1 (\%)}
    \label{Tab:MapBeliefsS1}
    \vspace{0.7cm}
\end{subtable}
\begin{subtable}{\textwidth}
    \centering
    \includegraphics[trim=0mm 0mm 0mm 0mm, clip, width=1.00\textwidth]{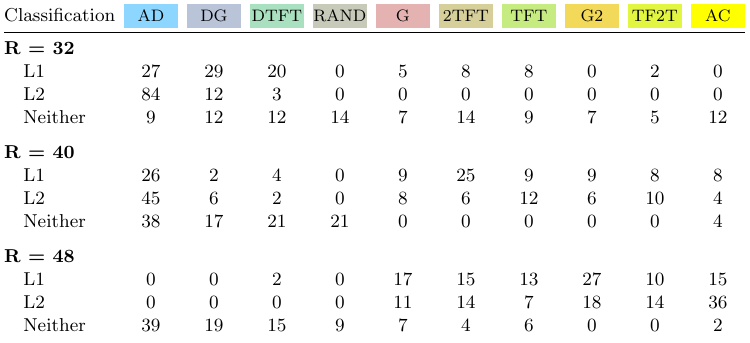}
    \vspace{-0.5cm}
    \caption{Strategy choices in Supergame 1 (\%)}
    \label{Tab:MapStrategiesS1}
      \vspace{0.6cm}
\end{subtable}
\begin{minipage}{0.95\textwidth}
\footnotesize
\noindent
Notes: $R$ is the return to joint cooperation, varied between subjects (see Section \ref{Sect:Design}). We classify subjects using the wider classification described in the main text.
For each treatment-category pair, we report the mean belief weight placed on each strategy (panel (a)) and the percentage of subjects choosing each strategy (panel (b)).\end{minipage}
      \vspace{0.4cm}
\caption{Empirical beliefs and strategies by treatment-category pair in Supergame 1}
\label{Tab:MapS1}
\end{table}

\clearpage

\section{Robustness: Choice of error tolerance}\label{Sect:RobustMore}

Our notion of balance-optimal error tolerance maximizes the share of subjects classified uniquely as level-1 or level-2, and therefore strikes a balance by tolerating enough error for the level-$k$ model to explain observed behavior, without tolerating so much error that the behavior of different types overlaps too much. A natural question remains: are our results sensitive to using the balance-optimal error tolerance?

We answer this question in two ways. First, we show that our results are robust when we allow for sampling variability by bootstrapping the balance-optimal error tolerance and then using error tolerances toward the ends of the range of bootstrapped values. In particular, Table \ref{Tab:Regressions_RobBoth} shows that the regression results in Table \ref{Tab:Regressions} are robust when we use error tolerances at the 5th percentile or 95th percentile of the distribution of bootstrapped balance-optimal error tolerances (the table notes provide details). Cognitive ability continues to predict higher level-$k$ reasoning, and the level of reasoning continues to increase with experience. Specifically, the positive effects of cognitive ability and experience on the probability of being classified as L2 continue to be statistically significant, with $p<0.01$ in all cases.\footnote{Compared to Table \ref{Tab:Regressions}, the effect of cognitive ability on the probability of being classified as L2 in Cols. 1-2 of Table \ref{Tab:Regressions_RobBoth} is larger at the 5th percentile and smaller at the 95th percentile (although effects are similar in Cols. 5-6). We note that higher error tolerance admits more noise relative to the exact level-$k$ predictions, potentially making cognitive ability less predictive for the L1 vs. L2 classification.} Similarly to Table \ref{Tab:Regressions}, we continue to find little systematic evidence of an effect of personality traits. Thus, our conclusions do not hinge on the exact value of the balance-optimal error tolerance.

Second, we show that our results are robust when we do not use the balance-optimal error tolerance at all. In particular, Table \ref{Tab:RegressionsWide_Rob} shows that the regression results in Table \ref{Tab:RegressionsWide} are robust when we use the wider classification methodology from Section \ref{Sect:Robustness}, but with $\eta=1$ instead of the balance-optimal error tolerance (the notes to Table \ref{Tab:RegressionsWide_Rob} provide details). At $\eta=1$, any amount of noise is tolerated, and thus we classify all subjects as L1 or L2. We find that the positive effects of cognitive ability and experience on the probability of being classified as L2 continue to be statistically significant, with $p<0.05$ in all cases.\footnote{Compared to Table \ref{Tab:RegressionsWide}, the effect of cognitive ability on the probability of being classified as L2 is smaller in magnitude: again, as noted above, higher error tolerance admits more noise in level-$k$ behavior, potentially making cognitive ability less predictive.}

\newpage

\begin{table}[ht]
\vspace{-0.8cm}
\centering
\begin{subtable}{1.0\textwidth}
\includegraphics[width=0.95\textwidth]{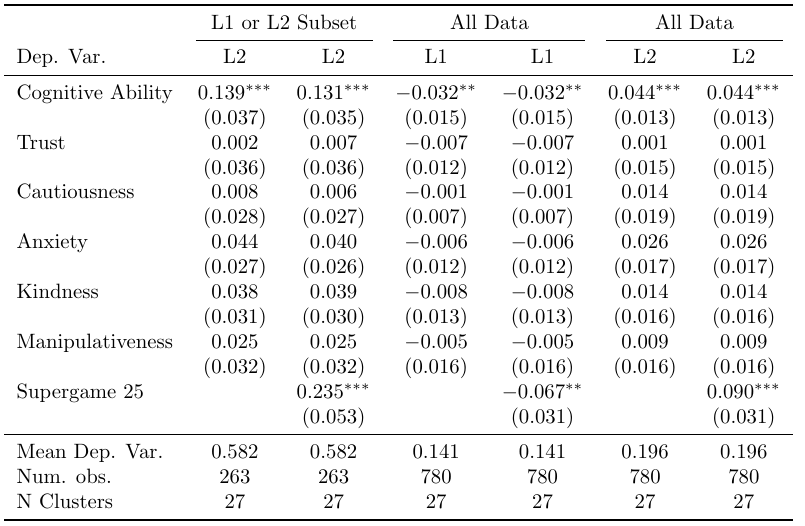}
\vspace{-0.1cm}
\caption{At 5th percentile of distribution of bootstrapped balance-optimal error tolerances}
\label{Tab:Regressions_Rob1}
\vspace{0.5cm}
\end{subtable}
\begin{subtable}{1.0\textwidth}
\includegraphics[width=0.95\textwidth]{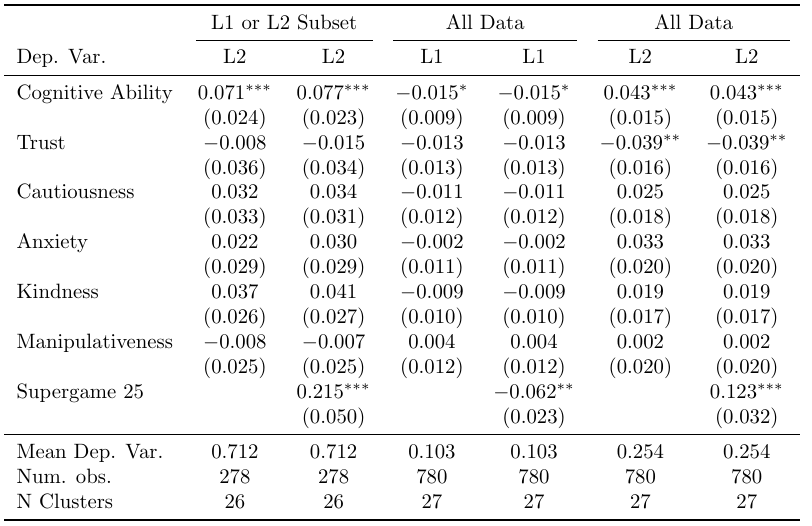}
\vspace{-0.1cm}
\caption{At 95th percentile of distribution of bootstrapped balance-optimal error tolerances}
\label{Tab:Regressions_Rob2}
\end{subtable}
\begin{minipage}{0.92\textwidth}
\vspace{0.5cm}
\footnotesize
\noindent
Notes: See the notes to Table \ref{Tab:Regressions}. Here, we run the same regressions, except that we replace balance-optimal error tolerance values with percentiles from the distribution of bootstrapped balance-optimal error tolerances. In panel (a), we replace $\eta^*_1=0.113$ and $\eta^*_{25}=0.118$ with 0.098 and 0.100, the values at the 5th percentile for Supergame 1 and 25, respectively. In panel (b), we replace $\eta^*_1=0.113$ and $\eta^*_{25}=0.118$ with 0.128 and 0.124, the values at the 95th percentile for Supergame 1 and 25, respectively. We bootstrap at the session level, stratified by treatment (value of $R$), with 10,000 replications.
\end{minipage}
\vspace{0.2cm}
\caption{Effects of cognitive ability, personality, and experience on level-$k$:\\ Robustness using bootstrapped balance-optimal error tolerance}
\label{Tab:Regressions_RobBoth}
\end{table}

\clearpage

\begin{table}[ht]
\vspace{0.4cm}
\centering
\includegraphics[trim=0mm 0mm 0mm 0mm, width=0.95\textwidth]{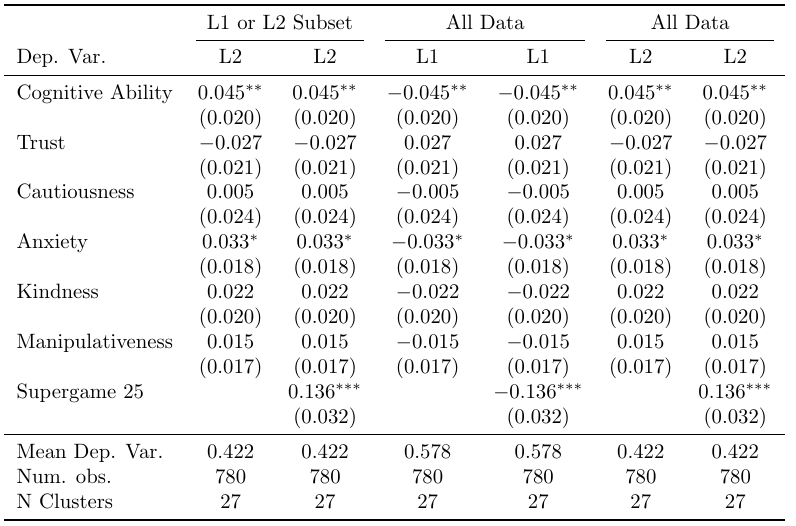}
\begin{minipage}{0.93\textwidth}
\vspace{0.3cm}
\footnotesize
\noindent
Notes: See the notes to Table \ref{Tab:RegressionsWide}. Here, we run the same regressions, using the same wider classification of L1 and L2 methodology, but setting error tolerance at its maximum value of $\eta=1$ instead of at balance-optimal $\eta^*_1$ and $\eta^*_{25}$. Specifically, at $\eta=1$, all subjects are classified as Both. Consider a given subject in either Supergame 1 or 25: as we lower error tolerance below $\eta=1$, eventually we reach a value at which the subject's beliefs are consistent with one and only one of the level-1 and level-2 types (Assumptions \ref{Ass:1} and \ref{Ass:2} define belief consistency). Here, we assign the subject to L1 (L2) if her beliefs are consistent only with the level-1 (level-2) type at this lower $\eta$. Using this methodology, we are able to classify all subjects as L1 or L2. Therefore: (i) the L1-or-L2 subset coincides with the full sample, so Cols. 1-2 replicate Cols. 5-6; and (ii) L1 effects in Cols. 3-4 are the negatives of L2 effects in Cols. 5-6. However, we keep all six columns for consistency and comparison with earlier tables.
\end{minipage}
\vspace{0.3cm}
\caption{Effects of cognitive ability, personality, and experience on level-$k$:\\ Robustness without using balance-optimal error tolerance}
\label{Tab:RegressionsWide_Rob}
\end{table}

\clearpage

\section{Conclusion}\label{Sect:Conclusion}

This paper advances our understanding of boundedly rational behavior in repeated strategic interactions by introducing a novel model of level-$k$ reasoning that operates at the level of supergame strategies. Unlike existing approaches that analyze reasoning round by round, our framework captures how individuals form and act on beliefs about entire strategic plans over the course of a repeated game. Using experimental data from the Indefinitely Repeated Prisoner’s Dilemma, where both strategies and beliefs were elicited, we demonstrate that many subjects exhibit behavior consistent with level-1 or level-2 reasoning. Importantly, because the level-$k$ model is fundamentally belief-driven, the joint elicitation of strategies and beliefs facilitates identification of distinct reasoning processes. We find systematic variation in the depth of reasoning: individuals with greater cognitive ability and more experience are more likely to be classified as higher level types. These findings contribute to the behavioral foundations of repeated game theory and highlight the importance of belief formation and individual heterogeneity in strategic environments that unfold over time.

Our work provides a foundation for several promising directions for future research. First, our model of level-$k$ reasoning could be applied to additional environments, including repeated coordination games, bargaining games, and market entry games. Second, the model could be extended to include more behavioral types, such as worldly players who best respond to the empirical distribution of behavior or types that incorporate social preferences. Third, the model of level-$k$ reasoning could incorporate explicit belief updating, in order to investigate how players learn about the distribution of types over time and how this learning relates to individual characteristics. Finally, incorporating heterogeneous cognitive costs or bounded memory would enable richer predictions about when and why players adopt different levels of reasoning in repeated games.

\clearpage

\renewcommand{\refname}{References}
\fontsize{11}{10}
\begin{spacing}{1.36}
\interlinepenalty=10000
\bibliographystyle{experiment}
\bibliography{GR_Level_k}
\end{spacing}

\clearpage

\thispagestyle{empty}

\text{}
\vspace{3cm}
\begin{center}

\textbf{\Huge{Appendix A:}}

\vspace{0.5cm}

\textbf{\Huge{Supplemental Appendix}}

\vspace{1cm}

\LARGE{(Intended for Online Publication)}

\end{center}

\clearpage

\setcounter{table}{0}
\renewcommand{\thetable}{A.\arabic{table}}

\setcounter{figure}{0}
\renewcommand{\thefigure}{A.\arabic{figure}}

\setcounter{section}{0}
\renewcommand{\thesection}{Appendix A.\arabic{section}}
\renewcommand{\thesubsection}{Appendix A.\arabic{section}.\arabic{subsection}}

\pagenumbering{arabic}
\renewcommand*{\thepage}{Appendix A, page \arabic{page}}

\section{Further figures and tables}\label{Sect:Further}

\begin{figure}[H]
  \centering
  \includegraphics[width=0.7\linewidth]{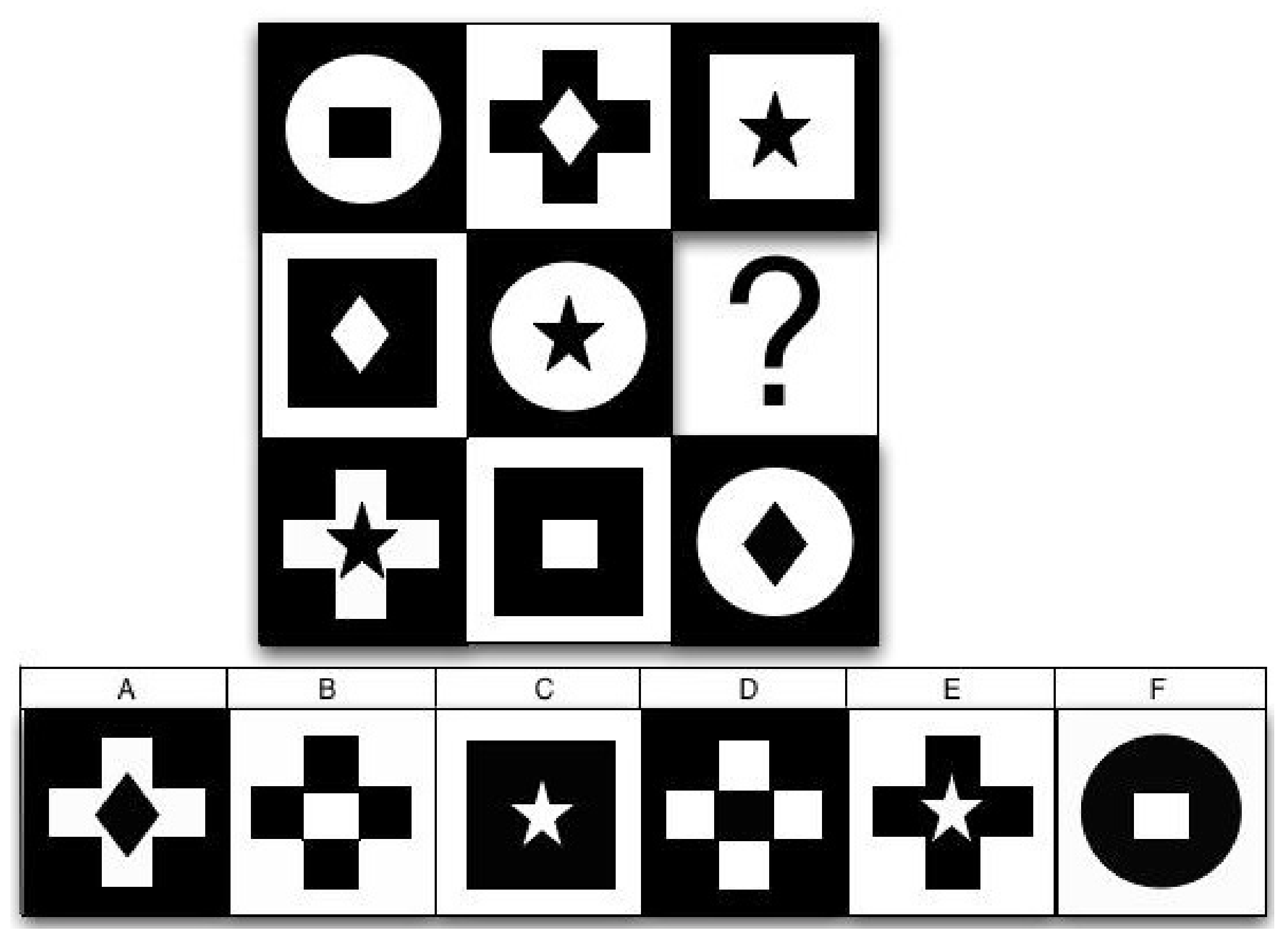}
  \caption{Sample item from test of matrix reasoning}
  \label{Fig:Matrix}
  \vspace{1cm}
\end{figure}

\begin{figure}[ht]
    \centering

    \begin{subfigure}{0.48\textwidth}
        \centering
        \includegraphics[width=\linewidth]{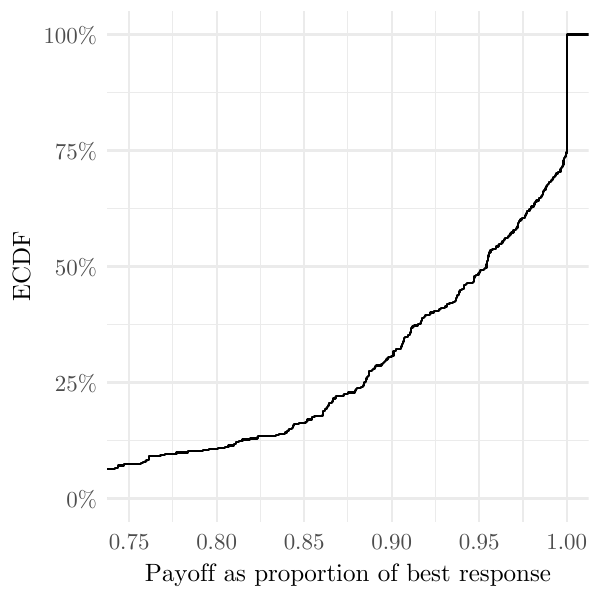}
        \caption{Supergame 1}
        \label{Fig:CDF1}
    \end{subfigure}
    \hfill
    \begin{subfigure}{0.48\textwidth}
        \centering
        \includegraphics[width=\linewidth]{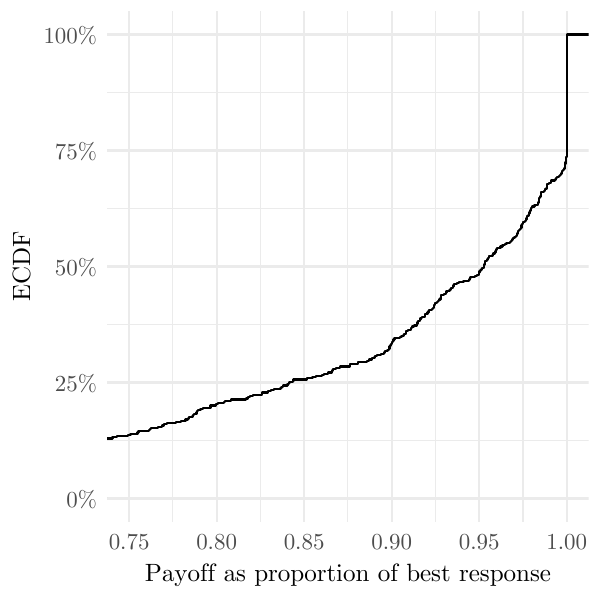} %
        \caption{Supergame 25}
        \label{Fig:CDF25}
    \end{subfigure}

    \vspace{0.5cm}

    \begin{minipage}{0.87\textwidth}
    \footnotesize
    \noindent
    Notes: The figure shows the empirical cumulative distribution function (ECDF) across subjects of the expected payoff of the subject’s chosen strategy, given her beliefs, as a proportion of the expected payoff of the best response to her beliefs. Before calculating expected payoffs, we normalize payoffs to range from 0 to 1.
    \end{minipage}

        \vspace{0.3cm}

    \caption{ECDF of payoffs as proportion of best response to beliefs}
    \label{Fig:CDF}
\end{figure}

\clearpage

\begin{table}[ht]
\centering
\includegraphics[
    trim=12mm 0mm 2mm 6.5mm,
    clip,
    width=1.03\textwidth]{Tables/10-belief_clusters_table_sup01.pdf}
\vspace{-0.3cm}
\caption{Belief clusters in Supergame 1 (all clusters)}
\label{Tab:BeliefsS1_All}
\end{table}

\begin{table}[ht]
\centering
\includegraphics[
    trim=12mm 0mm 2mm 6.5mm,
    clip,
    width=1.03\textwidth]{Tables/10-belief_clusters_table_sup25.pdf}
\vspace{-0.3cm}
\caption{Belief clusters in Supergame 25 (all clusters)}
\label{Tab:BeliefsS25_All}
\end{table}

\begin{table}[ht]
\centering
\begin{subtable}{\textwidth}
    \centering
    \includegraphics[trim=0mm 0mm 0mm 0mm, clip, width=1.00\textwidth]{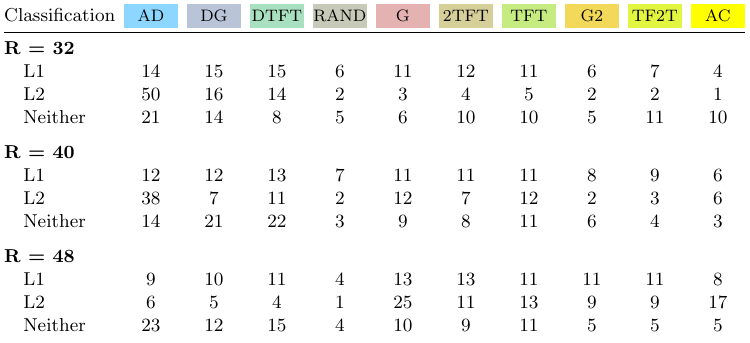}
       \vspace{-0.5cm}
    \caption{Beliefs in Supergame 25 (\%)}
    \label{Tab:MapBeliefsS25}
    \vspace{0.7cm}
\end{subtable}
\begin{subtable}{\textwidth}
    \centering
    \includegraphics[trim=0mm 0mm 0mm 0mm, clip, width=1.00\textwidth]{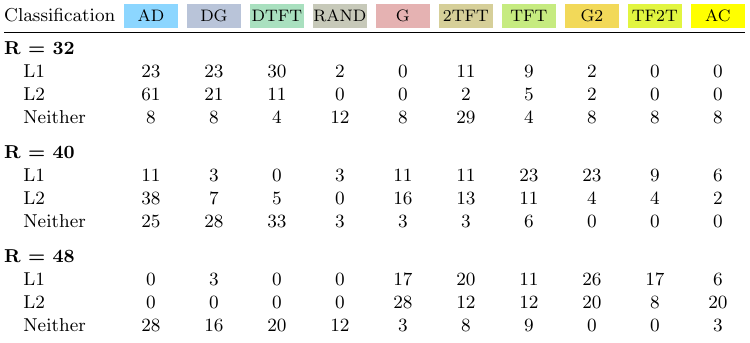}
    \vspace{-0.5cm}
    \caption{Strategy choices in Supergame 25 (\%)}
    \label{Tab:MapStrategiesS25}
      \vspace{0.6cm}
\end{subtable}
\begin{minipage}{0.95\textwidth}
\footnotesize
\noindent
Notes: See the notes to Table \ref{Tab:MapS1}: we use the same methodology, now for Supergame 25.\end{minipage}
      \vspace{0.4cm}
\caption{Empirical beliefs and strategies by treatment-category pair in Supergame 25}
\label{Tab:MapS25}
\end{table}

\begin{landscape}

\thispagestyle{empty}

\text{}
\vspace{3cm}
\begin{center}

\textbf{\Huge{Appendix B:}}

\vspace{0.5cm}

\textbf{\Huge{Experimental Screenshots}}

\vspace{1cm}

\LARGE{(Intended for Online Publication)}

\end{center}

\clearpage

\setcounter{table}{0}
\renewcommand{\thetable}{B.\arabic{table}}

\setcounter{figure}{0}
\renewcommand{\thefigure}{B.\arabic{figure}}

\setcounter{section}{0}
\renewcommand{\thesection}{Appendix B}

\pagenumbering{arabic}
\renewcommand*{\thepage}{\arabic{page}}

\renewcommand{\headrulewidth}{0pt}
 \fancypagestyle{landscape}{ %
        \fancyhf{}
    \fancyfoot[C]{%
            \vspace*{-15.7cm} %
            \makebox[\textwidth][c]{\hspace{19cm}\rotatebox{90}{\thesection, page \thepage}} %
        }
    }                        \pagestyle{landscape}

\section*{}
\refstepcounter{section}
\label{Screen:Match}

\begin{textblock*}{\textwidth}(1.0cm,2.0cm)
 \rotatebox{90}{
    \includegraphics[width=1.65\textwidth]{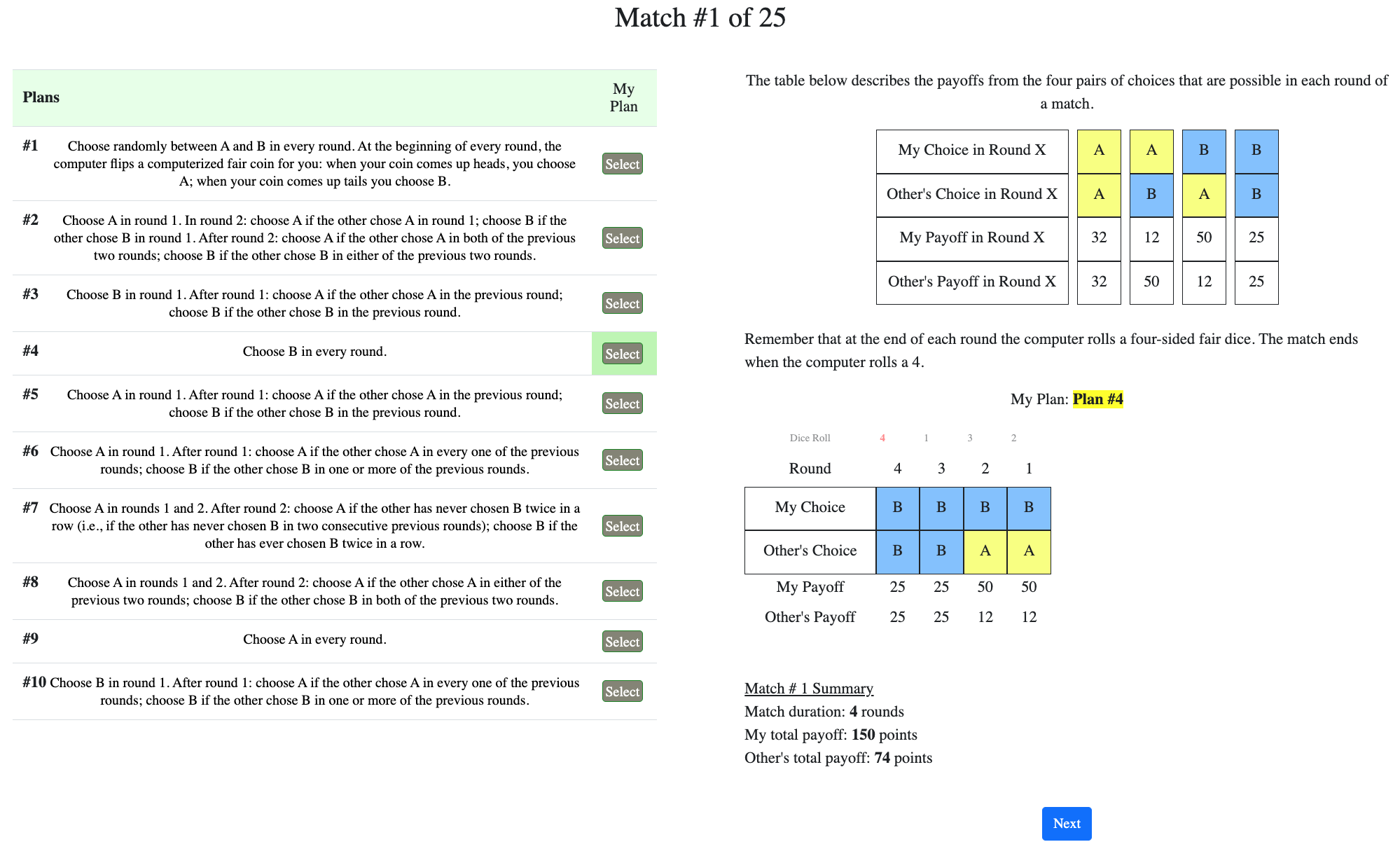}
    }
\end{textblock*}

\begin{textblock*}{\textwidth}(17.7cm, 2.5cm)
 \rotatebox{90}{
 \parbox{25.5cm}{
   \textit{Notes: The order that the ten strategies appeared on the subject’s screen was randomized across subjects. In this example, the order of the strategies is: RAND; 2TFT; DTFT; AD; TFT; G; G2; TF2T; AC; DG.}
    }}
\end{textblock*}

\clearpage

\section*{}
\label{Screen:Belief1}

\begin{textblock*}{\textwidth}(1.0cm,2.0cm)
 \rotatebox{90}{
    \includegraphics[width=1.65\textwidth]{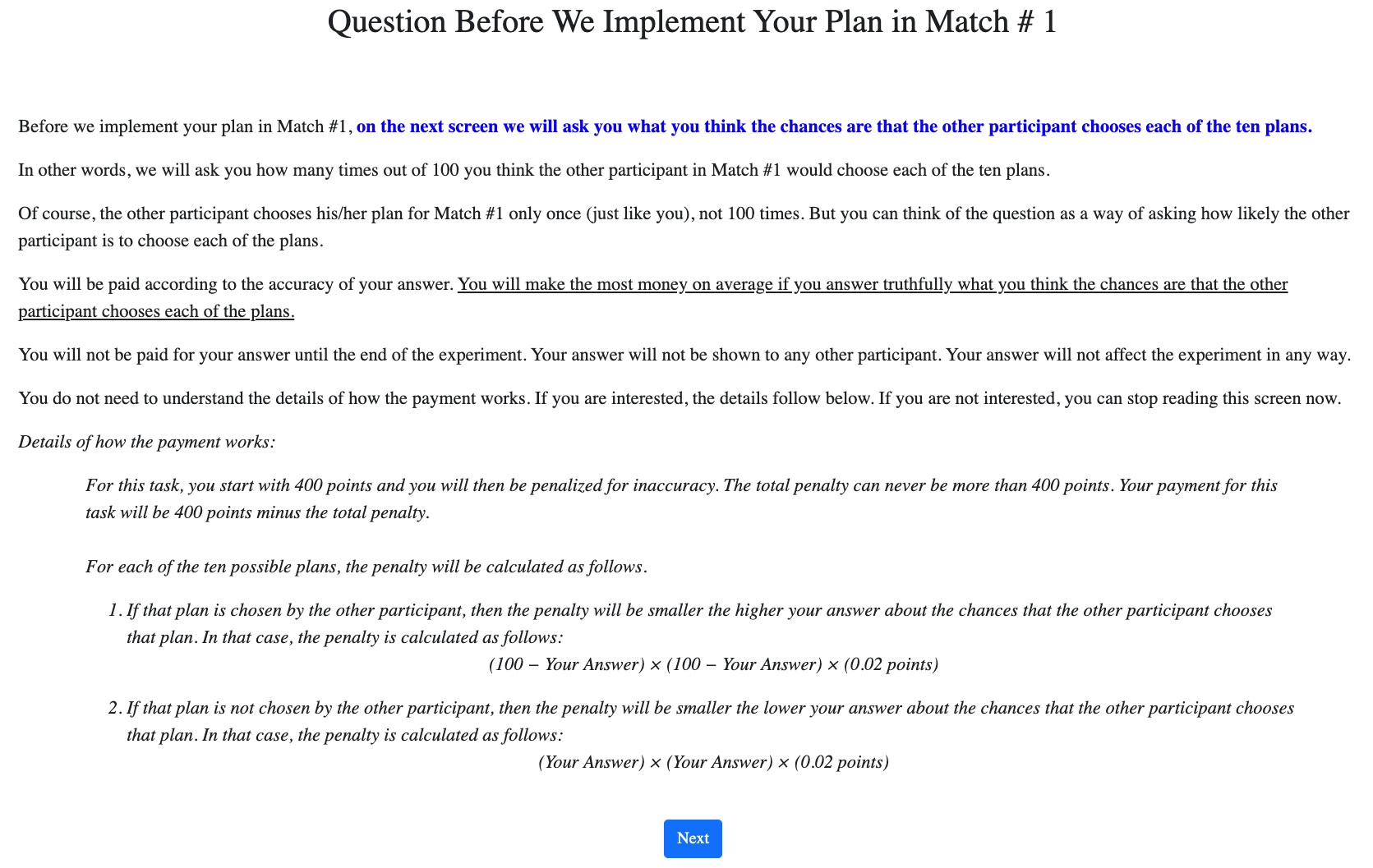}
    }
\end{textblock*}

\clearpage

\section*{}
\label{Screen:Belief2}

\begin{textblock*}{\textwidth}(1.0cm,2.0cm)
 \rotatebox{90}{
    \includegraphics[width=1.65\textwidth]{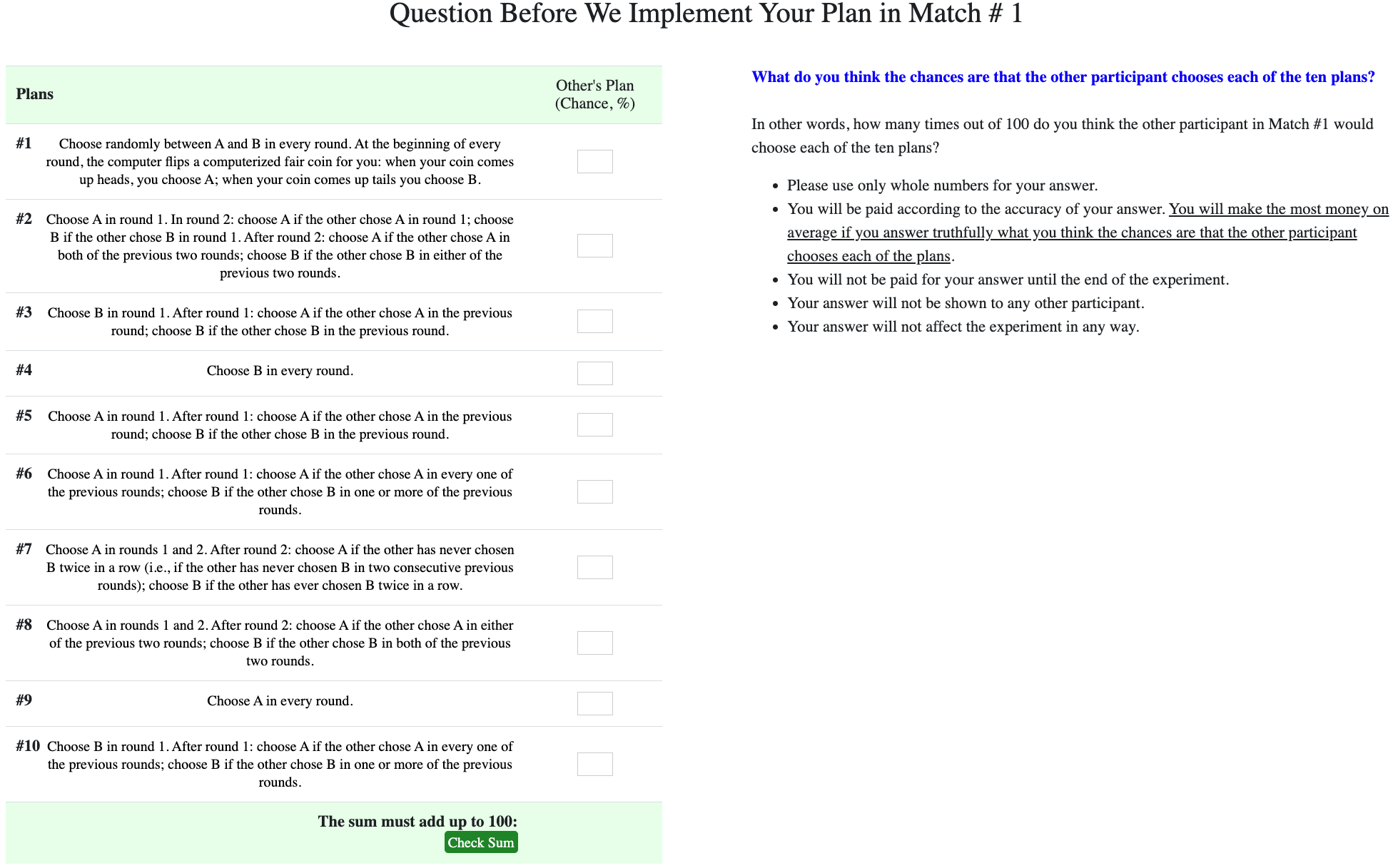}
    }
\end{textblock*}

\end{landscape}

\end{document}